\documentclass{eas}
\usepackage{graphicx}
\begin{document}

\title{Open clusters and associations in the Gaia era} 
\author{Estelle Moraux}\address{Univ. Grenoble Alpes, IPAG, F-38000
  Grenoble, France}\secondaddress{CNRS, IPAG, F-38000 Grenoble, France
  \\ \email{estelle.moraux@univ-grenoble-alpes.fr}}
%
%
\begin{abstract}
  Open clusters and associations are groups of young stars,
  respectively bound and unbound, that share the same origin and
  disperse over time into the galactic field. As such, their formation
  and evolution are the key to understand the origin and properties of
  galactic stellar populations. Moreover, since their members have
  about the same age, they are ideal laboratories to study the
  properties of young stars and constrain stellar evolution theories.

  In this contribution, I present our current knowledge on open
  clusters and associations. I focus on the methods used to derive the
  statistical properties (IMF, spatial distribution, IMF) of young
  stars and briefly discuss how they depend on the environment. I then
  describe how open clusters can be used as probes to investigate the
  structure, dynamics and chemical composition of the Milky Way. I
  conclude by presenting the Gaia mission and discuss how it will
  revolutionize this field of research.
\end{abstract}
\maketitle

\section{Introduction}



It is commonly said that stars form in ``clusters''. However it would
be more correct to say that stars form in groups with $N=10$ to $10^5$
and a disctinction should be made between young open clusters and
associations. While both of them correspond to physically associated
groups of population I stars, sharing the same origin and moving
together through the Galaxy, a cluster is gravitationally bound and an
association is not.

\begin{figure}[htbp]
  \centering
  \includegraphics[width=0.9\textwidth]{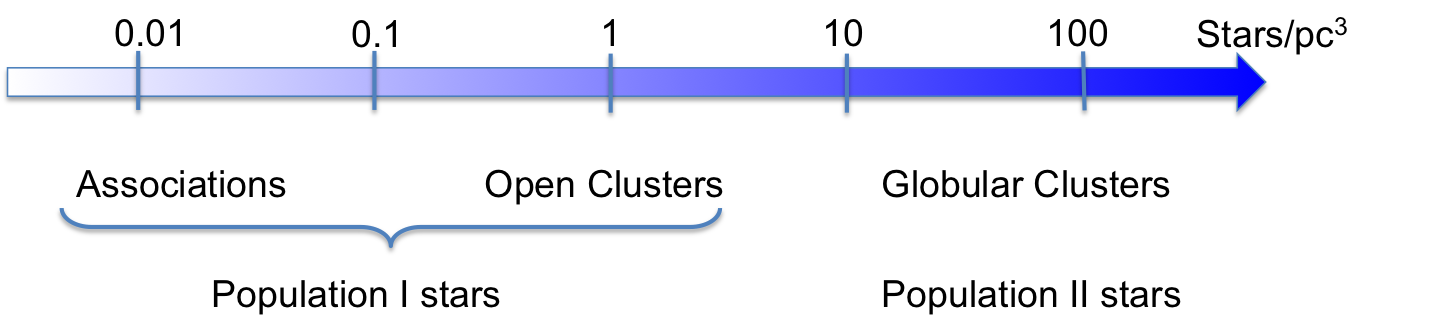}
  \caption{Typical density of associations, open clusters and globular
  clusters.}
  \label{density}
\end{figure}

Observationally, a cluster (like the Pleiades) shows a clear
concentration of young stars above the surrounding stellar background
while an association corresponds to a region where there is a higher
than normal density of O-B stars (OB associations, such as Orion OB1)
or T-Tauri stars (T associations, like TW Hydra). The average stellar
density of open clusters can go from 0.1 to about 10 stars/pc$^3$ in a
typical diameter of a few parsecs, yielding a few tens to 10$^5$
stars. An association contains typically 10 to 1000 stars which are
spread over a much wider area (several tens to a few hundred parsecs
in diameter), corresponding to a stellar density of $\sim0.01$
stars/pc$^3$ or less (see Figure~\ref{density}).

Another difference between a cluster and an association arising
directly from their definition is the lifetime. An open cluster is
loosely bound and its star members will stay together for about 100
Myr to a few Gyr for the densest ones, before it gets disrupted by the
galactic tidal field. On the contrary, an association is unbound and
disperses rapidly, over a timescale of 10 to 100 Myr.

This means that it is much more difficult to detect associations and
get a full census of their population. Very large areas of the sky
need to be covered to identify a few members among thousands of
unrelated field stars based on colors or proper motions. In
particular, Hipparcos data have been very powerful in detecting OB
association (e.g. \cite{deZ99}). However, caution must be taken when
using kinematic informations for $\sim100$ Myr associations as lots of
field objects may share the same proper motion due to resonant trap in
the galactic field (e.g. ABDor; \cite{fam08}). To date, about 80 O-B
associations and 3000 open clusters are known in the solar
neighborhood (up to $\sim1.8$ kpc; \cite{kha13}),
but they may be up to $10^5$ in the Galaxy.


Despite the above differences, open clusters and associations show
strong similarities. They are located in the arms of the Milky Way and
other spiral galaxies, they are composed of young stars that recently
formed in the disks of galaxies, and they continue to form (as opposed
to globular clusters). All the stars belonging to the same
cluster/association have the same age (with maybe a spread of a few
Myr) and metallicity, having formed from the same cloud of gas and
dust. Open clusters and associations can therefore be used to probe
the galactic disk structure and formation rate, as well as to study
the young star properties and their formation process. Where does the
galactic disk population come from: open cluster, association, or both
? Is there any prefered mode of star formation ?  Do clusters and
associations form similarly but evolve differently or do they form
from different physical conditions (density, turbulence, magnetic
field...) ? How does the cluster/association environment affect the
stellar properties ?


In order to be able to answer the above questions, we first need to be
able to distinguish a cluster from an association, or in other words
to know whether a stellar group is gravitationally bound or not. The
total energy of a system is $E=T+U$ with $U=-GM_c^2/R_g$ the potential
energy, and $\displaystyle T=\frac{1}{2}M_c\sigma_v^2$ the kinetic
energy. $M_c$ is the cluster mass, $R_g$ the {\it gravitational}
radius and $\sigma_v$ the 3D velocity dispersion. A system is bound if
$E<0$, i.e., if $\displaystyle \sigma_v<\sqrt{2\frac{GM_c}{R_g}}$ or
$\sigma_v<\sqrt{2}\,\sigma_v^{vir}$ where $\sigma_v^{vir}$ corresponds
to the {\it virialized} velocity dispersion given by
$2T+U=0$. Therefore a subvirial or virialized system is bound while a
supervirial system can be marginally bound or unbound.

In practice, it is not such an easy task to know the dynamical state
of a stellar group as one first needs to estimate its mass,
gravitational radius, and 3D velocity dispersion. Masses are estimated
either directly by summing up individual masses, or indirectly from
the tidal radius. Indeed $\displaystyle M_c=\frac{4A(A-B)r_t^3}{G}$
(\cite{kin62}) where A and B are the Oort constants and $r_t$ is the
tidal radius. $R_g$ is usually derived from the core radius $r_c$
($\sim0.2-0.3 R_g$) that is obtained by fitting a King distribution to
the radial profile -- but this can only be done if the group is
centrally concentrated. Typical values for open cluster core radii are
1-2 pc and 10-25 pc for tidal radii.  The velocity dispersions in open
clusters are typically only a few km/s, indicating that individual
stellar velocities have to be measured with an accuracy $<1$km/s which
requires high resolution spectroscopic observations.  Moreover
$\sigma_v$ has to be corrected from bias introduced by binaries
(e.g. \cite{chb14}, \cite{cot12}). Recent studies have shown that
among the young massive star clusters Westerlund 1 is subvirial and
strongly bound (\cite{cot12}) while NGC3603 is about virialized and
marginally bound (\cite{pan13}). The Cyg OB2 association is indeed
unbound (\cite{kim07,kim08}; \cite{wri14}) but
does not show any coherent expansion pattern on large scale. And the
nearby star forming region IC348 is supervirial but probably bound
thanks to the gas mass (\cite{cot15}).

For faraway regions that are barely resolved and for which $\sigma_v$
cannot be measured, another diagnosis has been proposed by
\cite{gpz11}. Noticing that all the stellar systems that stayed
together for more than 30Myr and are most probably bound have an
age that is larger than their crossing time $t_{cr}$, they suggest to
use the ratio $\pi=\textrm{age}/t_{cr}$ as a proxy to distinguish
between clusters ($\pi>1$) and associations ($\pi<1$). However the
$\pi$ value is only indicative for regions younger than 10Myr and the
distinction is not so clear.

For embedded regions, in which a lot of gas remains in particular, the
differentiation between cluster and association is meaningless. Indeed
a group that is bound thanks to the gas mass may become unbound after
gas removal.

In this lecture, I will first present the properties of young stars
in clusters (section~\ref{stars}), before describing how the global
properties of open clusters can be used as probes of the galactic thin
disk in section~\ref{global}. I will then present the Gaia mission and
discuss what we will learn from it  in section~\ref{gaia}.


\section{Properties of stars in young clusters/associations}
\label{stars}

Young clusters and associations harbour homogeneous populations and
represent ideal laboratories to study the properties of young stars
and brown dwarfs in various environments with different ages and
physical properties (e.g. density and/or metallicity). 

\subsection{Census of the population}

\subsubsection{Membership criteria}
In order to identify bona fide cluster/association members from
observations, several criteria can be used. First of all, since they
share the same age, metallicity and distance (for clusters only - for
associations, individual parallax measurements are mandatory to take
each star distance into account), their location in the HR diagram
defines an isochrone whose color and luminosity can be used to
distinguish members from field interlopers. Deep large scale
photometric surveys in the optical, and then in the near-infrared to
look for low mass objects and/or in extincted regions, have been
designed over the last twenty years for this purpose.

However, photometric data alone are not sufficient to distinguish
cluster members since older but closer field stars may share the same
properties. Complementary spectroscopic observations are thus often
used to determine the candidate spectral types and/or to look for
youth indicators such as the presence of lithium, a strong activity
level or a low surface gravity.

Another powerful diagnosis is to look at kinematics (proper motion and
radial velocity). Indeed, cluster or association members share the
same space motion, with a velocity dispersion of the order of a few
km/s , which offers an additional criterion to identify them. So far
proper motion measurements from ground wide field imaging with a
reasonable time baseline ($<10$ years) was only possible for fast moving
($>10$ mas/yr) groups. In some cases, like in Orion, the cluster momentum
is roughly aligned with the line of sight and radial velocity obtained
from high resolution spectroscopy may be the only way to disentangle
young overlapping associations (e.g. OB1a, OB1b and 25 Ori;
\cite{jef06}, \cite{bri07}, \cite{max08}, \cite{sac08}).

\cite{bou13} recently developped a new astrometric pipeline able to
combine and analyse archival data from various cameras at different
telescopes to get a proper motion measurement accuracy better than 1
mas/yr (corresponding to 0.5 km/s at 100 pc). Thanks to Gaia, even
more accurate astrometric data will soon become available, which is
going to revolutionize our ability to derive membership from proper
motion (see section \ref{gaia}). Astrometric measurements
can also be used to determine the moving group convergent point
(point in the sky towards which the velocity of all group members
seems to point due to the projection on the celestial sphere) and
constrain membership (\cite{gal12}).

\subsubsection{Membership analysis}

Not so long ago, the usual way to search for members was to select
candidates based on their photometric properties and then to do
follow-up observations for each object to confirm its membership based
on spectroscopy and/or kinematics. Over the last ten years, full sky
proper motion catalogs such as PPMXL (\cite{roe10}) or UCAC4
(\cite{zac13}) combining large photometric surveys have been
available, and the use of astrometric data became more systematic to
assess membership. Following the work of \cite{san71}, probabilies to
belong to the group based on the are defined allowing to select only
the most probable members. However, the contrast between the cluster
population and the field decreases at fainter magnitudes as the proper
motion measurement accuracy gets lower. Thus a single probability cut
for all magnitudes is going to miss more low-mass members than
high-mass ones and the level of contamination and/or incompleteness
depends on luminosity. This has to be taken into account when studying
the cluster population properties, but it is not always easy to
estimate properly this bias.

New statitistical methods (e.g. \cite{sar14}, \cite{mal13},
\cite{riz11}) combining all the available informations (positions,
proper motions, colors, and magnitudes) in a multi-dimensional
analysis are being proposed to get self-consistent and robust
membership probability whatever the object brightness. These Bayesian
approaches can allow for a full treatment of uncertainties and missing
data and may easily be extended to include other observables (like
radial velocity, distance, rotation...), making obsolete the classical
membership selection methods.

This methodology revolution is the consequence of the ever larger
amount of more accurate and sensitive data with broader spatial,
temporal and wavelength coverage that is becoming available.  A good
way to illustrate this is to look at recent studies made on the
well-known Pleiades cluster. While Moraux et al. (\cite{mor03})
presented the results from a 6.4 square degrees imaging survey down to
$I\sim22$ mag (which was one of the deepest and largest study in the
Pleiades at that time), UKIDSS photometric data in five passbands as
well as proper motion measurements for about 1 million sources spread
over $\sim80$ deg$^2$ have been analyzed less than 10 years later
(\cite{lod12}). In the framework of the DANCe
project\footnote{http://project-dance.com}, \cite{bou13} even
increased the number of analyzed sources by a factor of about 3.5 and
derived proper motions with an unprecedented accuracy (better than 1
mas/yr down to $i\sim23$ mag) by combining visible and near-infrared
images obtained over 15 years from nine different
instruments. \cite{sar14} developped a new methodology to infer
membership probabilities from this multi-dimensional data set in a
consistent way, using a maximum-likelihood model using Bayes’
theorem. As a result, more than 2100 high-probability members have
been identified, among which $\sim800$ are new (\cite{bou15}). These
authors almost doubled the number of high probability members, and
multiplied by five the number of substellar candidates, even though
the Pleiades cluster had been intensively studied.

\subsection{IMF}

One of the main motivation to obtain a full census of a cluster
population is to derive its initial mass function (IMF) that
corresponds to the distribution of stars at birth as a function of
their mass, and to investigate its possible universality in a wide
variety of environments. The IMF is an important physical property as
its shape puts constraint on the star formation process and drives the
evolution of clusters and galaxies. Several reviews of excellent
quality have been published recently (Bastian et al. \cite{bcm10},
\cite{jef12}, \cite{kro13}, \cite{off14}) and I refer the reader to
these for more details.

\subsubsection{Usual functional forms}
\label{imf_form}

There are usually three main functional forms used to describe the
shape of the IMF:
\begin{itemize}
\item a power-law $\displaystyle \chi(m) = \frac{dN}{dm} \propto m^{-\alpha}$ (or
  $\displaystyle \phi(m) = \frac{dN}{dlog m} \propto m^{-\Gamma}$ with
  $\alpha=\Gamma+1$). This form has been first proposed by \cite{sal55}
  with $\alpha=2.35$ (or $\Gamma=1.35$) for masses larger than
  $1M_{\odot}$ and revised to a serie of three segmented power-laws
  (\cite{kro01}) to represent the full mass range;
\item a log-normal $\displaystyle \phi(m) \propto exp \left[-
    \frac{(log m - log m_c)^2}{2\sigma^2} \right]$ (\cite{ms79}). More
  recently, \cite{cha05} proposed a similar form up to $1M_{\odot}$
  but a Salpeter power-law at higher masses;
\item and a tapered power-law $\displaystyle \chi(m) \propto m^{-\alpha}
  \left[1-e^{(-m/m_p)^{-\beta}} \right]$ (\cite{dem05}).
\end{itemize}
But see also \cite{cw12}, \cite{mas13}, \cite{bas15} for alternatives.

\subsubsection{IMF derivation}

The methodology to compute the IMF of a resolved cluster consists in
1) determining the luminosity function (LF) of the sample of
high-probability members and correct it from contamination,
incompletesseness and extinction when necessary; 2) converting the LF
into a present day mass function (PDMF) using an evolutionary model;
and 3) correcting the PDMF for stellar evolution and/or dynamical
evolution to get the IMF.

The first step of this process is critical and great caution must be
taken when dealing with a sample of cluster candidate members. Indeed,
contamination of photometric surveys by field stars (dwarfs and
giants) and extragalactic objects (galaxies, quasars) is always
present whatever the method used for membership analysis. In
principle, astrometric data should allow to reject most of the
extragalactic sources as they are not moving but only if the mean
cluster momentum is much larger than the measurement
uncertainties. Incompleteness may also be an issue if the survey does
not extend beyond the cluster tidal radius, especially if there is
mass segregation as the incompleteness level will then depend on
mass. Moreover, objects might be missed around bright stars due to
contrast issue, in crowded regions or in area with high extinction.

As for the second step, this is probably the one that suffers from the
largest uncertainties. Indeed the conversion of colors and magnitudes
into masses depends on evolutionary models which may not be very well
constrained, especially at low masses and very young ages. As an
example, the ONC mass function may show very different features below
0.3$M_{\odot}$ when using different models (see
e.g. \cite{dar12}). Moreover, even for the same set of models, the
results may be very sensitive to the cluster estimated age or distance
as shown by \cite{sch13}. For example, using an age of 1 Myr instead
of 3 Myr for NGC1333 yields a different mass function. In addition to
these model issues, individual stellar properties such as rotation
rate and/or activity may influence the object luminosity and therefore
the mass estimate. Indeed fast rotators may have a larger radius which
would result in a larger luminosity and an overestimated mass. At the
opposite, strong magnetic activity could create large cool spots which
would reduce the effective temperature and underestimate the mass (see
e.g. \cite{moh09}, \cite{sta14}). The accretion history of a star
could also affect its luminosity (up to $\sim10$ Myr, see
\cite{bar09}), and therefore its mass derivation, depending on how the
star absorbs the accretion energy. And last but not least, multiple
systems that are not resolved in the photometric images (typically for
separations $<1$ arcsec, or $<100$ AU at 100 pc) are considered as single
objects and their mass is therefore erroneous. If the effective
temperature is used for conversion, then the obtained mass will be
close but smaller than the real primary mass, but if instead the
luminosity is used, the mass will be larger and actually closer to the
total mass of the system. This means that even if using the same set
of models, and the same age and distance, results may be different
depending on how the conversion of the photometric properties into
mass is performed.

\begin{figure}[htbp]
  \centering
  \includegraphics[width=0.55\textwidth]{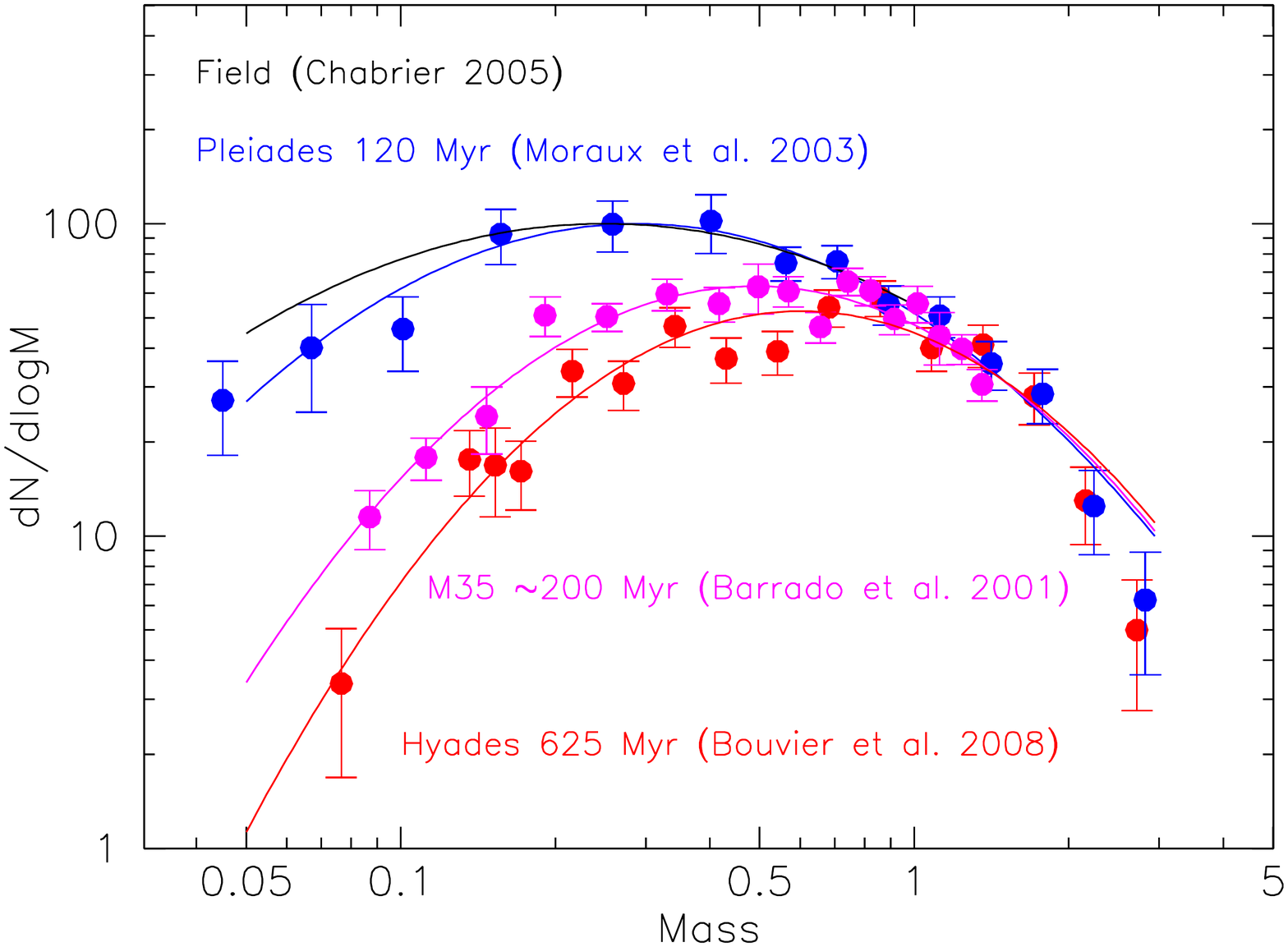}
  \includegraphics[width=0.44\textwidth]{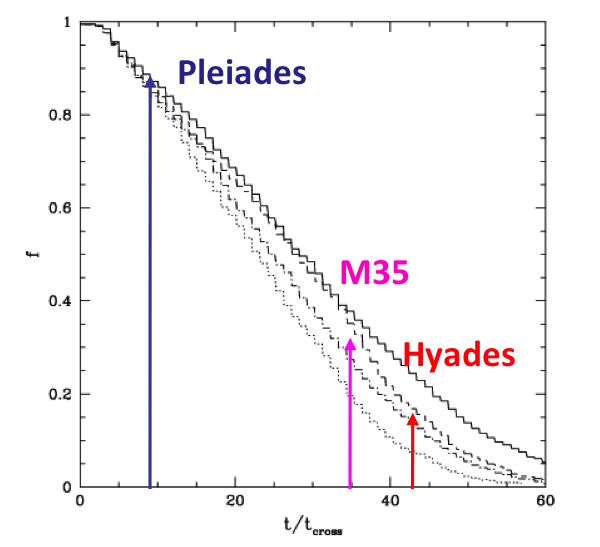}
  \caption{{\it Left:} Present day mass function of three open
    clusters: the Pleiades (blue), M35 (magenta), and the Hyades
    (red). The field mass function from \cite{cha05} is shown for
    comparison (black).  {\it Right:} Results from N-body simulations
    showing the fraction of remaining brown dwarfs in the cluster as a
    function of age in crossing time unit (adapted from
    \cite{ada02}).}
  \label{mf_evol}
\end{figure}

The third and last step to derive an IMF is to correct the present day
mass function from evolutionary effects. After a relaxation time
$t_{rlx} \sim (0.1N/lnN) R/\sigma_v \sim (0.1N/lnN) t_{cr}$ where $N$
is the total number of cluster members and $t_{cr}$ is the crossing
time, enough exchanges of energy via 2-body interaction between
cluster members have occurred to reach equipartition. This means in
particular that lower mass objects have a higher velocity dispersion
than more massive ones. As a consequence, mass segregation sets-up
(the relative number of low mass stars with respect to high mass stars
is smaller in the cluster center than in peripheric areas), and low
mass members are lost preferentially by the cluster as a larger
fraction of them have a velocity greater than the escape velocity.
The latter effect can be clearly seen from observations: the Hyades
PDMF (\cite{bou08}) shows a clear deficit of brown dwarfs and low mass
stars compared to that of the Pleiades, which is fully consistent with
the dynamical evolution predicted by N-body simulations (see
Figure~\ref{mf_evol}). Consequently, the peak of the MF shifts towards
higher masses as the cluster evolves, which has also been shown by
\cite{dem10}. For a typical open cluster with $R\sim4$ pc,
$\sigma_v\sim1$ km/s, $M\sim1000 M_{\odot}$ and $N\sim2000$, one has
$t_{cr}\sim4$ Myr and $t_{rlx}\sim 25 t_{cr} \sim 100$ Myr. However,
these values may change a lot from region to region. A denser cluster
will evolve faster (as $t_{cr}$ varies with $1/\sqrt{\rho}$ where
$\rho$ is the cluster mean mass density), but so will a low-N cluster
($t_{rlx} \sim 2 t_{cr}$ for $N\sim100$). It is important to keep in
mind that a 10 Myr small group may be already relaxed and may have a
PDMF that resembles that of a much older cluster. What really matters
here is the age of the region with respect to its relaxation time.

Another strong limitation in the IMF derivation comes from stellar
evolution that leads to the death of massive stars. Indeed the upper
mass function cannot be directly determined from stellar counts above
a given mass that depends on the age of the cluster, but has to rely
on model.

In practice, most of the published PDMF are not corrected from the
aforementioned evolutionary effects as it would add even more
uncertainties in the IMF determination. It is important however to be aware
of them and to look only at young and dynamically unevolved regions to
be sure that the derived mass function is representative of the IMF.

\subsubsection{IMF universality}

In their review, Bastian et al. (\cite{bcm10}) compiled a large number
of young cluster and association mass functions to look for any
possible variations of the IMF. Although each individual fit by a
tapered power-law (see \ref{imf_form}) yields different parameters,
they conclude that there is no evidence for variation when comparing
the IMF of Milky Way stellar clusters as the observed differences are
smaller than the uncertainties inherent to the mass function
derivation (as described above). \cite{dib14} made a thorough
comparison of the mass function of eight young galactic clusters using
Bayesian statistics. The IMF inter-cluster comparison shows that the
parameters distributions do not overlap within the $1\sigma$ level and
the author concludes that the IMF is not universal. However, although
the analyzed clusters have been chosen to reduce the uncertainties on
the mass function derivation (i.e., only complete data have been taken
into account and masses have been obtained using the same set of
evolutionary models), they are still likely underestimated. The
uncertainties on the cluster age and distance estimates are not taken
into account, and the effects of rotation, magnetic activity and
multiplicity are neglected. It is not possible to draw any conclusion
on IMF variations until the cluster parameters have been ascertained
and the models have been anchored at young ages and low
masses. Hopefully, the situation should change soon thanks to the
advent of Gaia.

\begin{figure}[htbp]
  \centering
  \includegraphics[width=0.8\textwidth]{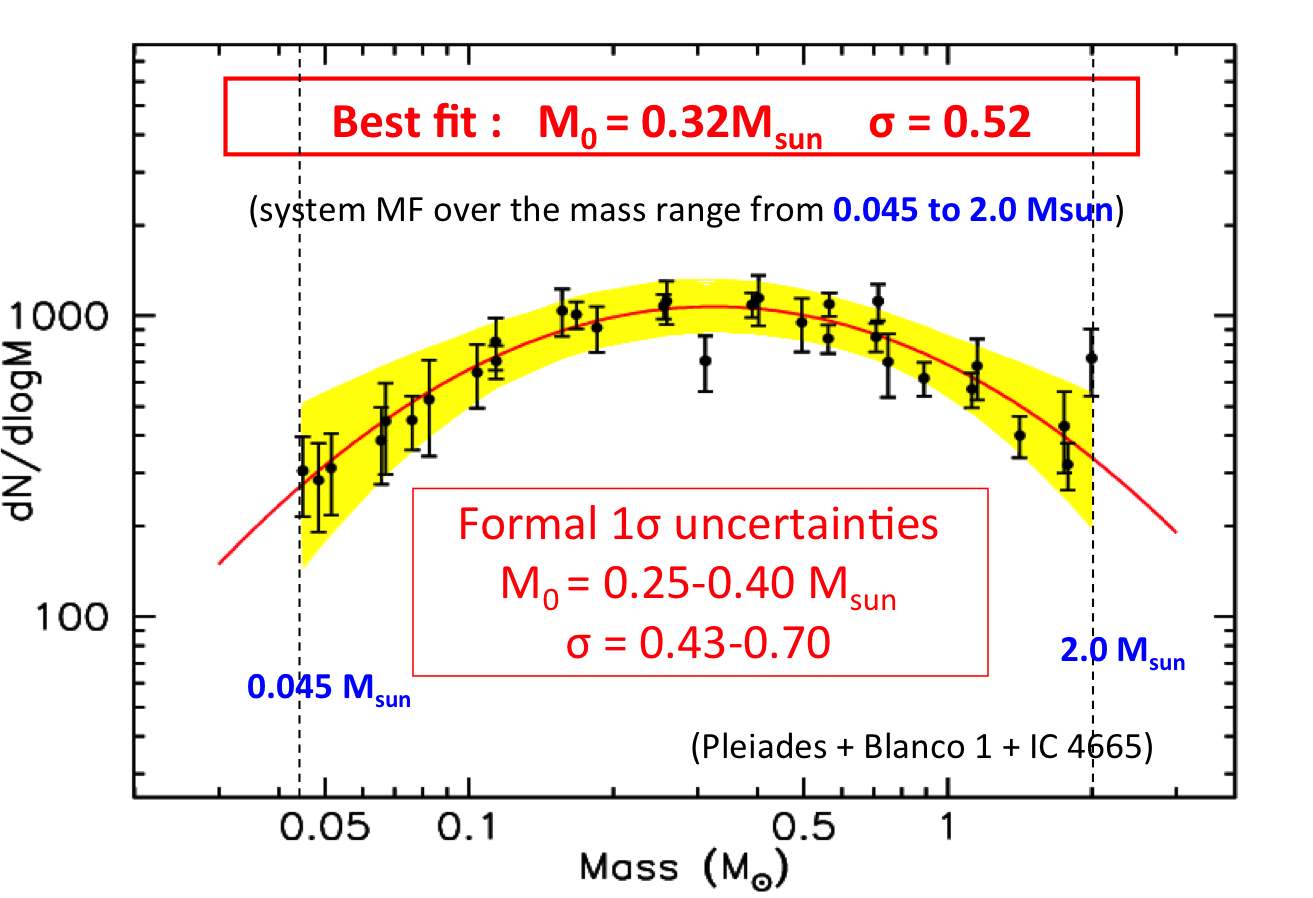}
  \caption{Superimposed system mass function of three open clusters:
    the Pleiades (Moraux et al. \cite{mor03}), Blanco~1 (\cite{mor07}), and IC4665
    (\cite{dew06}). The red curve corresponds to the best lognormal
    fit to the combined data in the mass range [0.045-2] $M_{\odot}$,
    and the yellow area shows the $1\sigma$ uncertainty of the fit.}
  \label{yoc_cfh}
\end{figure}

One of the best constrained mass function to date is probably the
Pleiades one since the cluster age ($125\pm8$ Myr, \cite{sta98}) and
distance ($130\pm10$ pc, \cite{val09}, \cite{mel14}) are known with an
accuracy better than 10\%. The cluster is rich enough to minimize
Poissonian errors and close enough to sample well into the substellar
domain. The best lognormal fit to the {\it system} mass function in
the $[0.03-2] M_{\odot}$ range gives $m_c=0.25 M_{\odot}$ and
$\sigma=0.52$ (Moraux et al. \cite{mor03}). Overlapping this fit to
the mass function of a large number of open clusters and star forming
regions reveals that they are all consistent within the uncertainties
over the same mass range (see \cite{off14}, their Figure 2). Moreover,
when considering three open clusters that have been analyzed in the
exact same way by the same group (from the data reduction to the mass
derivation and using data obtained with the same instrument and the
same evolutionary models), their mass functions overlap almost
perfectly and the best fit to the combined data gives
$m_c= 0.32^{+0.08}_{-0.07} M_{\odot}$ and
$\sigma= 0.52^{+0.18}_{-0.09}$ (see Figure~\ref{yoc_cfh}).  This
suggests that the cluster stellar IMF may be universal, at least from
$0.03 M_{\odot}$ to $2 M_{\odot}$, and that a lognormal form may be a
good representation in this mass range (see, however, \cite{luh03} for
Taurus).

Below $0.03 M_{\odot}$, the situation is less clear. There may be some
hint for variations as an excess of probable members between 0.01 and
0.03 $M_{\odot}$ has been claimed in Upper Sco (\cite{lod13}) and in
$\sigma$-Ori (\cite{pen12}). However, some issues remain regarding
these results as most of the candidates still need to be confirmed
spectroscopically and the mass-luminosity relation is very uncertain
at such low masses. Future instruments such as the JWST will help to
investigate the IMF in various regions at low masses and down to the
planetary mass regime, to eventually search for a cut-off in the lower
IMF.

Another way to look for IMF variations is to investigate very
different environments in terms of density or metallicty. Young
massive clusters in the Milky Way (see also the lecture from
N. Bastian for extra-galactic ones) offer the possibility to look at
the high mass part of the IMF which has been found to be flatter than
Salpeter in some cases.  This is in contradiction with recent estimate
of the galactic disk IMF that is found to be steeper than Salpeter
above $1M_{\odot}$ with $\alpha$ close to 3 (\cite{cze14}). While the
Arches mass function above $1M_{\odot}$ is found to be consistent with
a Salpeter IMF (\cite{hab13}), the Westerlund 1 one is controversial
since \cite{lim13} report a flatter IMF while \cite{gen11} report a
normal IMF. In the Quintuplet cluster, \cite{hus12} derived an IMF
that is flatter than Salpeter, but this may be due to mass segregation
as they only investigated the central part of the cluster. In the
young nuclear cluster in the Galactic center, \cite{lu13} found a MF
slope of $\alpha = 1.7 \pm 0.2$, which is again flatter than
Salpeter. In NGC3603, the mass function is also a power-law with
$\alpha \sim 1.7$ but from $0.4M_{\odot}$ to $10M_{\odot}$
(\cite{har08}, \cite{sto06}).  No turnover has been found in this
region yet, while there is one around $0.5M_{\odot}$ in Tr14
(\cite{roc11}) and $0.15M_{\odot}$ in Westerlund~1 (\cite{and16}). The
JWST and then the ELT will allow to refine these estimates and
determine the IMF down to substellar masses in these regions.

In low metallicity environments, such as globular clusters (see
C. Charbonnel's lecture, and \cite{dem07}) and the galactic bulge
(\cite{cal15}), the IMF is found to be consistent with the lognormal
shape proposed by \cite{cha05} but the mass range probed by these
studies is very narrow ($0.2-0.8 M_{\odot}$).

In conclusion, while the stellar IMF does not seem to vary
significantly between $0.03M_{\odot}$ and $2M_{\odot}$ from region to
region and appears to be independent of density, the situation is not
as clear at lower masses nor at higher masses. In particular,
departures from the Salpeter's IMF have been found in young massive
clusters and in the galactic disk but in the opposite direction. These
possible variations still have to be understood in the framework of
star formation.

\subsection{Multiplicity}
\label{multiplicity}

Another ubiquitous outcome of star formation is stellar
multiplicity. Characterizing the frequency and main characteristics of
multiple systems and their dependencies on environment is therefore a
powerful tool to probe this process. This could also help to correct
the shape of the IMF from multiplicity and to disentangle features
that may be due to multiplicity properties variations rather than to
the individual star mass function. Duch\^ene \& Kraus (\cite{dk13})
published a very complete review on stellar and substellar multiplicity in
the field, but also in clusters and associations, and I refer the
reader to it.

I will only focus here on the main difference between dense clusters
and loose associations, namely the frequency of visual companions
(separation between 10 to 3000 AU) that shows a clear dichotomy
between these two kind of environments (see Duch\^ene \& Kraus
\cite{dk13}, their Figure 4). The visual companion frequency observed
in dense regions ($>50$ stars/pc$^2$) is similar to that of field
stars, but only about half that of low-density regions ($<10$
stars/pc$^2$), whatever their age. For spectroscopic binaries however,
no significant difference is observed whatever the regions.

This difference of visual companion frequency does not necessary imply
that the initial multiplicity properties are environment-dependent
however.  Dynamical evolution may naturally explain both the “low
multiplicity” and “high multiplicity” tracks, in which open clusters
and loose associations represent extreme cases of highly effective and
ineffective binary disruption respectively. Numerical simulations show
that frequent interactions in dense clusters can reduce the initial
multiplicity frequency by a factor of 2-4 within 1 Myr
(e.g. \cite{mk12}). Current observations do not
probe multiplicity in an early enough evolutionary stage to allow us
to conclude on the universality of a high multiplicity frequency.

\subsection{Substructures}

It is clear from observations that open clusters and associations have
a very different aspect: the former show a clear concentration of
stars that can often be centrally defined, while the latter are spread
over a large sky area. However, this difference is not so clear at very
young ages. Indeed, Spitzer images have revealed that all embedded
regions (bound or unbound) are substructured. High stellar concentrations
are surrounded by a more extended population of young stars following
the filametary structure of the parent molecular cloud. \cite{gut11}
have shown that there is a strong correlation between the surface
density of young stellar objects and the column density of gas. Some
massive star forming regions exhibit large deviation from this
correlation however as large stellar clusters
are offset from high extinction area (e.g. CepOB2), but this may be
due to gas dispersal by massive stars and non-coevality.

Recent multi-object high resolution studies indicate that
substructures are also present in the velocity space of young clusters
(e.g. NGC2264, \cite{fur06}) and are probably imprints of the
formation history. \cite{jef14} identified two dynamically independent
populations in Gamma Velorum: a virialized and centrally condensed
group that could be a bound remnant of a dense cluster, and a
supervirial, scattered population, spread over several degrees,
older by 1-2 Myr, and further away by $\sim10$ pc.

Thus, being able to identify and characterize subgroups of young stars
may shed light on the star formation process and help to understand
the difference on the initial conditions that will lead to a cluster
or an association. \cite{cw04} defined a parameter $Q$ to evaluate the
degree of substructure that is based on the Minimum Spanning Tree
(MST) algorithm (\cite{gr69}). It is defined as the ratio of the mean
length of the tree branches to the normalized mean separation
between stars. If $Q>0.8$, the spatial distribution is smooth and
centrally concentrated, substructured otherwise. The definition of a
threshold for the MST branch lengths or the nearest-neighbor distances
can also be used to identify groups (e.g. \cite{gut09}), in which the
more massive stars tend to be towards the center.

\cite{par14} performed N-body simulations of the dynamical evolution
of substructured systems and found that subvirial initial conditions
result in a global collapse that leads to a bound, centrally
concentrated ($Q>0.8$) and mass segregated cluster within a crossing
time (with $t_{cross}\sim1$ Myr). In contrast, if the initial
conditions are supervirial, no global mixing occurs ($Q$ remains
small), the whole region expands and becomes an unbound
association. These results suggest that clusters and associations may
arise from the same star formation process but have a different
dynamical evolution depending on the dynamical state of the region
after gas removal.


\section{Clusters as probes of the Galactic thin disk}
\label{global}

Open clusters (hereafter OC) are powerful probes to characterise,
study and analyse a variety of aspects related to the structure,
composition, dynamics, formation and evolution of the Milky Way. Not
being exhaustive, these aspects include:
\begin{itemize}
\item the Glactic spatial scales (scale height of the disk(s), height
of the Sun above the Galactic plane, distance to the Galactic Centre) ;
\item the spiral structure (tracing spiral arms) ;
\item the kinematics and dynamics
(rotation of spiral pattern, Galactic rotation curve, velocity of the
local standard of rest, Oort constants, orbits of OCs and their
relation to cluster survival/disruption) ;
\item the Galactic formation and evolution (age of the oldest clusters,
  abundance pattern).
\end{itemize}

The main advantages of using open clusters is that they can be found
at all ages and in a wide range of locations in the disk.  Their
parameters (age, distance, reddening) are available by, e.g.,
isochrone fitting of their colour-magnitude diagrams. Powerful
statistical methods (e.g. \cite{nj06}, \cite{mon10}, \cite{dia12})
have been developped yielding typical parameter
uncertainties of the order of 5-10\%.

There are also some disadvantages, or at least difficulties and
systematics in studying the Galaxy on the basis of OCs. Confusion due
to crowded backgrounds in the Galactic plane and dissolution of OCs
introduce biases in the identification of clusters. These effects
produce in particular selection effects against the low-mass OC that
are faint and poorly populated, but also against old clusters, which
are potentially poorer (due to dynamical evolution) than their younger
counterparts. The extent of such a selection is far from being well
known, but it must be kept in mind when deriving conclusions from the
observed global properties of open clusters. Moreover,interstellar
extinction affects the detectability of sources, introducing patterns
in the distribution of open clusters that are often overlooked.

\subsection{Catalogues}

Using OCs for studying the Galaxy relies on a huge amount of data
gathering and requires a catalogue of OC parameters (distances, ages,
velocities, metallicities, diameters). Thousands of studies of
individual OCs exist in the literature, often with nonconcordant
results, mainly due to the variety of techniques to reduce data and
derive parameters. This problem is well known and has motivated
several attempts to derive homogeneous parameter sets.

In the 1980s, the Lund catalogue (\cite{lyn82}) contained around 1200
clusters, among which a few hundred had some of their parameters
determined. A review on OCs and Galactic structure has been published
by \cite{ja82} and updated by \cite{jan88}.  The on-line WEBDA OC
database (formerly BDA; \cite{mer92}) appeared in the 1990s. More than
a list of clusters with their parameters (essentially from the Lund
catalogue), it included also actual measurements for their stars
(photometry, astrometry, spectral types, etc.).  From the
observational point of view, CCDs and infrared arrays became of common
use, and Hipparcos provided a wealth of astrometric and kinematic data
and settled the distance scale through accurate parallaxes up to
$\sim$ 100 pc.  Although the new data did not bring significant
results regarding the spatial structure of the Milky Way, the number
of abundance measurements and discovery of older OCs shifted the focus
on the origin and chemical evolution of the Galaxy. The main
review on the subject is that of \cite{fri95} and its updates.  The
2000s brought the \cite{dia02} catalogue (DAML02) of optically
reaveled OCs, which updated the Lund catalogue.

All-sky proper-motion and photometric surveys have enabled the
identification of OCs beyond simple visual recognition and opened the
possibility to derive homogeneous data sets of Galactic clusters.
Systematic searches for concentrations of stars at a common distance
or with common proper motion, cluster sequences in CMDs or simply
spatial concentrations of stars obeying some photometric criteria have
allowed the discovery of several hundred optically visible OCs
(e.g., \cite{pla98}; \cite{ale03}; \cite{kha05}). Lists of infrared
OCs and candidates have been compiled by \cite{bic03}, \cite{dut03},
Froebrich et al.  (\cite{fro07}, \cite{fro10}), and \cite{glu10} based
on 2MASS searches. These objects, detected in the
near-IR, added about 700 clusters to the optically identified
family. Taking advantage of more data becoming available, refined and
homogeneous determinations of spatial, structural, kinematic, and
astrophysical parameters for poorly studied clusters have been
published (e.g. \cite{bon06}; \cite{pis06}; \cite{kha09},
\cite{buk11}; \cite{tad11}; \cite{kha13}; \cite{dia14}), and a
number of clusters shown to be nonexistent have been removed from
the catalogues. It is important to mention that distinguishing
between true gravitationally bound, physical associations from chance
aggregates on the sky requires in-depth follow-up for individual
cluster candidates.  The latest version (3.5) of the DAML02
catalogue\footnote{The catalogue can be found on
  http://www.wilton.unifei.edu.br/ocdb/} now includes $\sim2200$
clusters. 99.7\% of them have estimates of their apparent diameters,
and 92.4\% have distance, reddening and age determinations. Concerning
the kinematic data, 97.2\% have their mean proper motions listed,
and 43.1\% their mean radial velocities. In total, 41.6\% have
distance, age, proper motion and radial velocity determinations
simultaneously. Including IR discoveries, Kharchenko et al. (2013)
published a list of $\sim2800$ open clusters called the Milky Way Star
Clusters (MWSC) sample. They found that it is almost complete up to a
distance of about 1.8 kpc from the Sun, except for the subset of the
oldest open clusters (log $t>9$), where there is evidence for
incompleteness within 1 kpc from the Sun (\cite{sch14}).

As for the total number of OCs in the Milky Way, estimates depend on
many hypotheses, such as the variation of the clusters volume density
with the distance from the Galactic Centre, or the cluster disruption
rate.  Recent studies indicate numbers on the order of 10$^5$ OCs
(\cite{bon06}; \cite{pis06}).

\subsection{Spatial distribution}

Understanding the true shape of the Milky Way is a challenging task
which depends essentially on the availability of complete Galactic
surveys of the spatial distribution of stars, star clusters,
molecular clouds, etc. The Milky Way is composed of the center and
nuclear bulge, the thin disk hosting most of the OCs, the thick disk
containing the old OCs, and the extended halo hosting old stars,
white dwarfs and globular clusters (e.g. \cite{maj93}). The gas and
stellar density of the disk is usually expressed as a combination of
exponential-decay profiles for the horizontal and vertical directions,
$\rho(r,z) \propto e^{(-r/R_D)} e^{(-|z|/z_h)}$ ,where $R_D$ is the
scale length and $z_h$ the scale height (e.g. \cite{bt87}).

Although the number of known OCs is small compared to stars, it is
relatively simple and accurate to derive distance and age for these
objects. In this sense, OCs may serve as a direct probe of the disk
structure. Analysing a sample of 72 OCs \cite{jp94} found that the
clusters younger than the Hyades (age $<$ 700 Myr) are strongly
concentrated to the Galactic plane. They are distributed almost
symmetrically around the Sun with a scale height perpendicular to the
Galactic plane of $z_h \sim 55$ pc. In contrast, the old OCs are more
spread with $z_h \sim 375$ pc.

The increased samples of OCs analysed in recent studies essentially
recover this result, although in a clearer way. In particular, it is
found that the disk scale height defined by the young clusters is
smaller than that derived from older clusters, with the young and
intermediate-age OCs ( $<$ 200 Myr versus $<$ 1 Gyr) having scale
heights of 48 pc and $\sim$150 pc, respectively (\cite{bon06}). For
older clusters, the vertical distribution is essentially uniform, so
that no scale height can be derived. The average scale height,
considering clusters of all ages, is $z_h \sim 57 \pm3$ pc
(\cite{bon06}; \cite{pis06}). Also, the distribution of OCs younger
than $\sim$1 Gyr clearly shows that $z_h$ increases with
Galactocentric distance, being about twice as large in regions outside
the Solar circle than inside it (\cite{bon06}).

Assuming that OCs are symmetrically distributed above and below the
plane, the zero point of the height distribution gives a measure of
the distance of the Sun from the Galactic plane. The current values
determined from OCs range from about 15 to 22 pc (\cite{bon06};
\cite{pis06}; \cite{jos07}; \cite{bf14}).

The volume density of clusters in their plane of symmetry is 800-1000
kpc$^{-3}$ and the surface density is around 100-120 kpc$^{-2}$ for
clusters closer than 1.8kpc (\cite{kha13}). The distribution
of the surface density as a function of distance is almost flat, with
a few exceptions.  A significant excess of the youngest clusters at
about 400 pc from the Sun is related to the Orion star formation
complex. The surface density of the oldest clusters ($>$1 Gyr)
presents a deficit below $\sim1.1$ kpc from the Sun, which may be
due to incompleteness effects.

The interpretation for the above OC scale heights to be smaller than
those attributed to the thin disk ($z_h \sim 0.6$ kpc) as derived by
means of stars (e.g.\cite{deb97}; \cite{alt04}; \cite{kae05}) is not
completely clear yet. This could be seen as a disk evolutionary
effect, by which the disk would thicken with age due to some heating
mechanism, or it could be an effect of the selective destruction of
OCs that tend to be close to the plane. From the perspective of OCs
seen as progenitors of the field population, \cite{kro02} has shown
that stars escaping the clusters, mainly during the process of gas
dispersal, can thicken Galactic disks and even produce thick disks,
which could also explain the scale height difference.

As for the the thicker scale height outside the solar circle, its
precise meaning is not straightforward and is most likely a
combination of effects such as the flaring of the disk and the
amplitude of the warp.

\subsection{Spiral arms}

\begin{figure}[htbp]
  \centering
  \includegraphics[width=0.7\textwidth]{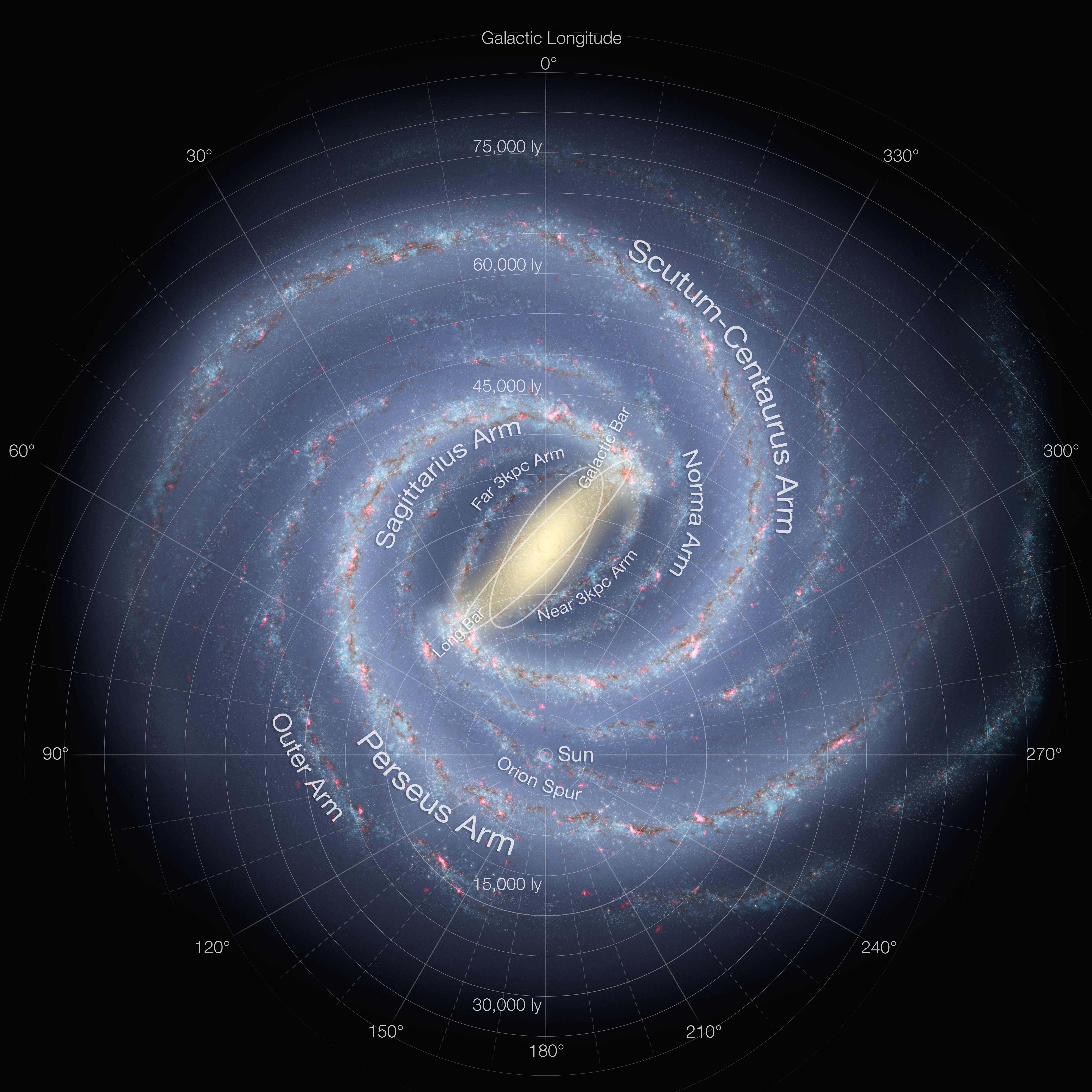}
  \caption{Artist's view of the Milky Way (Credits: R. Hurt,
    NASA/JPL-Caltech/SSC)}
  \label{MW}
\end{figure}

Although the spiral arm structure (see Fig.~\ref{MW}) in the Galactic
disk is an important component when studying the morphology and
dynamics of the Milky Way, a complete picture describing the
nature, origin, and evolution of this structure is still lacking. As
recently suggested by \cite{ben08}, there may be two major spiral
arms (Scutum-Centaurus and Perseus) with higher stellar densities and
two minor arms (Sagittarius and Norma) mainly filled with gas and star
forming regions. The nature of the Orion spur (or Local arm), i.e., the
spiral feature inside which our Sun is located, has been elusive to
present days. It might be a real arm, like Perseus, or a kind of
inter-arm feature, like the ones seen in many external spirals.

Very young open clusters ($<$12 Myr) and OB stars are powerful spiral
arm tracers, and can probe spiral features in very remote regions of
the Milky Way. At increasing cluster ages ($12<$age$<20$ Myr),
however, the spatial distributions get more scattered and no structure
can be detected, since these clusters had a lot of time to move away
from their birth-places (see \cite{dl05}).

Historically, star clusters have been used to map the spiral structure
of the Milky Way mainly in the third Galactic quadrant ($180<l<270$
deg ). In this area the extinction is small (\cite{moi01}) because the
young disk is significantly warped (\cite{moi06}), and star clusters
can be detected to very distant regions (\cite{car10}).  A major
breakthrough has been reported in \cite{vaz08}, where distances of a
large sample of young open clusters (mostly from \cite{moi01}) are
compared with CO clumps all the way to 20 kpc from the Galactic center
in the anticenter direction.

This study indicates that the Perseus arm does not seem to continue in
the third quadrant. Its structure is broken by the Local arm, which
extends all the way to the outer, Norma Cygnus arm. If this is
confirmed, it implies that the Local arm is a much larger structure
than currently believed, and behaves like a bridge, connecting the
Norma Cygnus arm in the outer disk possibly all the way to the
Sagittarius arm in the inner first quadrant of the Galaxy. This
picture of the Local arm as traced by young open clusters is in
contrast with the GLIMPSE realization of the third quadrant, but in
good agreement, e.g., with the HI study from \cite{lev06} who
found a conspicuous Hydrogen structure departing from roughly the Sun
location and entering the third quadrant, i.e., breaking Perseus and
reaching the outer arm.  It is therefore difficult to conceive that
the Orion spur is a spiral arm like, e.g., Carina-Sagittarius or
Scutum-Crux.

More recently, however, \cite{xu13} studied a sample of about 30
masers located in the Orion arm, and concluded that the kinematic of
the sample is typical of a grand design spiral arm, and that the Orion
spur is an arm of the same nature as Perseus or Scutum. \cite{rei14}
published a substantial compilation of over 100 high-precision
trigonometric distances to high-mass star forming regions using maser
VLBI observations. Interestingly, they found that the Perseus arm,
thought to be one of the major spiral arms of the Milky Way, has
little massive star formation over a 6 kpc long arc between Galactic
longitudes of 50 and 80 deg (\cite{cho14}; \cite{zha13}).  In
contrast, the Local (Orion) arm has comparable massive star
formation to its adjacent Sagittarius and Perseus arms.


\subsection{Kinematics}

The 1990s and 2000s witnessed a first explosion of astrometric data.
The main results being the Hipparcos and Tycho-2 catalogues (ESA 1997;
\cite{hog00}), the UCAC2, 3 and 4 compilations, and the PPXML
catalogue. However, while open clusters have often been used to
investigate the local behaviour of the Galactic rotation curve, a
proper rotation curve covering a sizable range of the Galaxy is still
lacking. The outcome of most kinematic studies has been to improve the
determination of parameters such as the local standard of rest (LSR)
rotation speed corresponding to the circular orbital speed, the Oort’s
constants and the velocity of the Sun (e.g. \cite{pis06};
\cite{sch10}; \cite{bru11}; \cite{bov12}; Reid et al. 2014).  The
rotation curve is found to be flat between $R_{GC}\sim5$ and 15
kpc. The fundamental parameters of the Galaxy, including the distance
of the Sun to the Galactic center, $R_0$, the circular orbital speed
at the Sun, $\Theta_0$, and the solar motions with respect to the LSR,
have been reevaluated to $R_0\sim8.3$ kpc, $\Theta_0\sim239$ km/s,
$U_{\odot}\sim11.1$ km/s, $V_\odot\sim12.2$ km/s and $W_\odot\sim7.2$
km/s. These values are quite different from the IAU standards, which
could have strong impact on our understanding of the Galaxy. For
example, increasing $\Theta_0$ by 20 km/s would result in a 30\%
increase in the estimate of the mass of the Milky Way (dark-matter
dominated).

In addition to these calculations, OCs have also been used recently to
determine the rotation speed of the spiral pattern.  Dias \& L\'epine
(2005) confirmed that spiral arms rotate as rigid bodies, and
determined the pattern speed by direct integration of cluster orbits
and by comparing the positions of clusters in different age groups,
leading to an adopted value of 24 km/s/kpc. As an additional result,
they found that the corotation radius is ($1.06 \pm 0. 08$)$\times$ the
solar Galactocentric distance.  It is also interesting to note that
\cite{bob07} found a different speed for the Orion arm compared to
Carina-Sagittarius and Perseus.  Another basic parameter is the
epicyclic frequency (the circular velocity obtained after subtracting
the Galactic rotation curve) which \cite{lep08} found to be $42\pm 4$
km/s/kpc at the solar radius.

\subsection{Metallicity distribution}

Open clusters, found at all ages and throughout the Galactic disk,
serve as excellent tracers of the overall chemical enrichment of the
disk. As they preserve the abundances of the gas from which they
formed, studying the chemical profiles of clusters of different ages
and locations provides the history of star formation and
nucleosynthesis throughout the galaxy and across its lifetime. Recent
high precision analyses indicate that stellar abundances among cluster
members are highly uniform, consistent with no intrinsic scatter
within the cluster (\cite{des06}), so one can be assured that the
determined abundances reflect the initial composition in a well-mixed
gas cloud.

The study of the radial metallicity distribution through open clusters
sees the first pioneering work at the end of the 70s, when
\cite{jan79}, using the photometric metallicity determination of about
40 clusters located within 6 kpc from the Sun, derived for the first
time, a radial gradient of [Fe/H]. They found that metallicity
decreases from the inner to the outer parts of the Galaxy at a rate
d[Fe/H]/dR$=-0.05$ dex/kpc. During the following decades, many authors
have tackled the argument, using gradually more accurate determination
of the cluster parameters, and, especially, of their metallicity
through low, medium and finally high resolution spectroscopy. A
non-exhaustive list of works dedicated to the study of the Galactic
metallicity gradient with open clusters includes \cite{fri95},
\cite{fri02}, \cite{yon05}, Sestito et al.  (\cite{ses06},
\cite{ses08}), \cite{pan10}, \cite{yon12}, \cite{hei14}. The most
common results are linear fits over the complete investigated
galactocentric radius ($R_{GC}$) range or gradients with different
slopes for inner and outer areas. While the inner disk ($<10$ kpc)
shows a steep gradient, the metallicity distribution of the outer area
is almost flat. \cite{twa97}, or more recently \cite{lep11}, provided
however another interpretation of the radial metallicity distribution
with two shallow plateaus and a step of $\sim$ 0.3 dex at
$R_{GC}\sim 9$ kpc. \cite{lep11} explained this as a consequence of
the corotation ring-shaped gap in the density of gas and that the
metallicity evolved independently on both sides.

Numerous studies have been devoted to understanding how the evolution
of the Milky Way shaped this chemical gradient (e.g. \cite{chi01};
\cite{ces07}; \cite{mag09}; L\'epine et al. 2011). Various physical
processes are responsible for the observed distribution of elements in
the Milky Way disk.  Chemical evolution models explore the influence
of star formation rates, recycling of material processed by the stars,
infall rates of extragalactic material onto the disk, and the
variation of those quantities with Galactocentric radius to determine
the combination that best reproduces the shape of the gradient at
various epochs. Recent works based on chemodynamical simulations
(\cite{ros08}; \cite{min13}, \cite{min14}) take into account stellar
migration, and the fact that the present-day Galactocentric radius of
a tracer may be different from its birth radius.

Large samples including clusters of different ages allowed to deal
with the temporal variation of the radial gradient (e.g.,
\cite{che03}; \cite{mag09}; \cite{jac11}; \cite{and11};
\cite{fri13}). \cite{net16} found that the older clusters ($\sim 1.7$
Gyr) appear more metal rich by about 0.07 dex than young clusters
($<0.5$ Gyr). This is clearly the opposite of what was commonly looked
for in the past, i.e., a decline of metallicity with cluster's age, but
it might be an observational confirmation of radial
migration. However, to further validate this result, a homogeneous age
scale for the clusters and a somewhat larger sample of older systems
are needed.

Taking all together, the literature results can
be summarised as:
\begin{itemize}
\item the global , i.e.,  including clusters of all ages, radial
  metallicity distribution traced by open clusters shows a steep inner
  gradient of -0.09 to -0.20 dex/kpc out to $R_{GC} \sim$10-12 kpc,
  and then presents either a shallow gradient or a plateau at $\sim$
  -0.3 to -0.5 dex from the breaking radius to the most faraway
  clusters, i.e.,  $R_{GC} \sim$ 20-23 kpc;
\item there is a notable dispersion in metallicity at any
  Galactocentric distance, typically of $\sim$ 0.5 dex;
\item concerning the time evolution of the gradient, there is a
  suggestion that younger clusters follow a flatter gradient than
  older clusters up to the breaking radius.  However, due to the small
  number of clusters in different age bins, the slopes of the gradient
  at different ages are sensitive to the binning and to the definition
  of the position of the breaking radius (cf., e.g., \cite{mag09},
  \cite{jac11}).
\end{itemize}

While overall metallicity and Fe abundances are useful in tracing the
general trends of chemical enrichment, the details of star formation
and nucleosynthesis are revealed in the behavior of elements such as
oxygen and the $\alpha$-elements Mg, Si, Ca, and Ti, and their
specific abundance pattern. These elements are formed through stellar
nucleosynthetic processes in massive stars, and, as a result, are
quickly recycled to the interstellar medium through mass loss and Type
II supernovae explosions. Elevated abundances of these elements,
relative to the solar ratios, point to episodes of rapid star
formation and distinctive star formation histories. 

Several groups (e.g., \cite{bt06} and the Bologna Open Cluster
Chemical Evolution project; Sestito et al.  \cite{ses08};
\cite{car07}; \cite{jac09}; \cite{fri10}; Yong et al. 2005) have led
the effort in characterizing the clusters chemical patterns.  Although
limited by small statistics, results indicate that groups of clusters
located at similar $R_{GC}$ share similar chemical patterns in
general, but remarkable differences are seen for some abundance ratios
in some clusters (\cite{mag14}).

\subsection{Age distribution}

The distribution of cluster ages sheds light on the balance between
cluster formation and disruption. If the cluster population seen
today were the result of a uniform formation rate combined
with an exponentially declining dissolution rate, one would expect to
see a simple exponential distribution with a characteristic single
lifetime.  What is seen is much different. The age distribution of
clusters shows three distinct populations with different timescales of
longevity.

The youngest cluster population has a  lifetime
of only a few tens of millions of years or less. A group of
such young clusters is apparent in any large sample, and their numbers rapidly
decrease with age beyond $\sim10$ Myr. These systems are stellar associations
that are not gravitationally bound, and are in the process of
dissolving and dispersing into the general field
population. Dissolution here occurs mainly because the gravitational
potential can be rapidly reduced by the impulsive gas removal by
supernovae explosions and massive star winds associated with this early
period. Thus, a significant fraction of the stars end up moving faster
than the scaled-down escape velocity, and may escape into the field
(e.g., \cite{gb06}).

Most of the clusters have a characteristic lifetime of a few hundred
million years and represent a rather homogeneous group. This
population dominates the Galactic system of open clusters and largely
defines the properties of the “typical” open cluster. The exponential
decay time for these clusters has been found to be between $sim$120
and 320 Myr (e.g., \cite{bon06}; \cite{pis06}). The cluster disruption
timescale varies with mass as $t_{dis} \sim M^{0.62}$
(e.g. \cite{lg06}), which means that solar neighbourhood clusters with
mass in the range $10^2-10^3 M_\odot$ dissolve on a timescale
$75< t_{dis} < 300$ Myr. This occurs because OCs continually undergo
mass segregation and evaporation, tidal interactions with Galactic
substructures, shocks with giant molecular clouds, as well as
mass-loss due to stellar evolution. By decreasing the total cluster
mass and the collective gravitational potential, these processes
affect the internal dynamics and accelerate the cluster dynamical
evolution until dissolution in the Galactic stellar field.

Nevertheless, as was first apparent in \cite{jan88} and reinforced
with larger samples in \cite{jp94}, the age distribution shows a long
tail to larger ages, in a distribution that cannot be fitted by a
single exponential with a decay time of a few hundred million
years. This third group represents only a few percent of the total
cluster population and includes members with ages approaching the age
of the Galactic disk, that can be characterized with lifetimes of 3-4
Gyrs. \cite{jp94} fit this older population in the cluster age
distribution with an exponential of 3-5 Gyrs, while Bonatto et
al. (2006) derived an exponential timescale of $2.4\pm1$ Gyr for this
old population. In either case, there is a substantial population of
old clusters not explained by the simple picture of uniform formation
rate combined with dissolution.

The mix and relative numbers of these populations vary with location
in the Galaxy, leading to the spatial distribution correlating with
age. The outer disk clusters are systematically longer lived and survive about twice as
long as a typical open cluster in the inner disk. The characteristic lifetime of open
clusters of all types seems to be a distinct function of galactocentric radius
(\cite{frie13}).

\subsection{Cluster mass function}

Although being a basic parameter for the study of cluster evolution, the
total mass $M_c$ of an open cluster can not be derived directly from the
observations. Indirect methods require, however, a number of
assumptions that can cause different biases in the resulting mass
estimates.

There are at least three independent methods for estimating cluster
masses, each with advantages and shortcomings.  The simplest and most
straightforward way is to count cluster members and to sum up their
masses. This is however challenging as it requires a complete census
of cluster members from the low mass objects to the highest mass
stars, and from the central parts to the outskirt of the
clusters. Incompleteness may arise from either the limiting magnitude
or the limited field of view or both. Moreover the conversion
from luminosity to mass relies on evolutionary models that may not be
very well constrained, especially a very young ages.
 
Another independent method is based on the virial theorem: the mass of
a cluster is determined from the 3-D stellar velocity dispersion $\sigma_v$ using the
relation $M_c=R_g \sigma_v^2 / G$, where $R_g$ is the cluster gravitational
radius. This does not require the membership
determination of all cluster members but the above relation assumes that the cluster
is virialized which may not be the case, especially for young
cluster. Moreover, the gravitational radius $R_g$ is obtained by 
fitting the cluster density profile (which may not be spherical
however), and may be underestimated if mass
segregation is present. The estimate of $\sigma_v$ is not an easy task
either as the typical velocity dispersion of open clusters is $\sim1$
km/s, and may be biased by the presence of binaries.
 
The third method uses the interpretation of the tidal interaction of a
cluster with the parent galaxy, and requires knowledge of the tidal
radius $r_t$ of a cluster. It gives the so-called ``tidal mass''
$\displaystyle M_c=\frac{4A(A-B)r_t^3}{G}$ (\cite{kin62}) where A and
B are the Oort constants. The tidal radius is usually obtained from a
three-parameter fit of a King profile to the cluster member surface
density distribution. This method assumes systems with spherical
density distributions whereas open clusters can have irregular or
elliptical shape. Another possible bias in the resulting tidal radius
may arise from the magnitude limit and/or the restricted covered area
used to build the cluster star sample from which the density profile
is built. This effect is particularly important for distant or mass
segregated clusters, for which the tidal radius may easily be wrong
by a factor of 2, and the mass by a factor of 8 as it varies with $r_t$
cubed. Tidal masses are, on average, larger than masses obtained
exclusively from star counts.

Whatever the method, the cluster present day mass function (CPDMF),
i.e., the mass distribution of Galactic open clusters, is
found to extend from $\sim$100 to $10^6 M_\odot$ and the high-mass
part of the CPDMF ($>10^3 M_\odot$) follows a power-law
$dN/dM_c \propto M_c^{-\alpha}$ with $\alpha \sim 2.0-2.2$. This
exponent increases with age as a direct consequence of the mass-loss
of clusters due to dynamical evolution. When taking into account only
the youngest clusters with age $<10$ Myr, $\alpha$ is found to be
around $1.7-2.0$ (e.g. \cite{pis08}). This result is
consistent with the embedded-cluster mass function that follows a
power law with a similar exponent (e.g. \cite{ll03}).

\subsection{Star formation rate in clusters}

Looking at the cluster age distribution, it appears that there has
been a local starbust between 220 and 600 Myr and that the star
formation rate (SFR) in the solar neighborhood is not constant. The
SFR is found to be around 4800 $M_\odot$/Myr during this period while
it is around half for the rest of the time (\cite{bb11}). There seems
also to be an excess of young clusters ($<9$ Myr), and the obtained
SFR in these young clusters is similar to the one derived for embedded
clusters, indicating that a majority of stars ($>80-90\%$) form in
groups.  However, many of these systems may not be bound and could
instead be associations in the process of dissolving.  Indeed, only a
fraction of $\sim 10\%$ survive the first 10 Myr (\cite{ll03})
suggesting that only a minority of field stars were born in groups
that were massive enough to become open clusters.

However, recent results on the multiplicity properties of field stars
(Duch\^ene \& Kraus \cite{dk13}), suggest that most of them cannot
come from loose associations either, as the observed visual companion
fraction (separation between 10 to 3000 AU, see
section~\ref{multiplicity}) is much lower in the field compared to
these regions. Therefore, most of the thin disk population seems to
originate from intermediate regions, that are dense enough to process
visual binaries but not massive enough to survive the first 10 Myr
after gas removal.


\section{What will we learn from Gaia?}
\label{gaia}

\subsection{The Gaia mission}

The Gaia satellite\footnote{http://www.cosmos.esa.int/gaia} is a
European Space Agency (ESA) cornerstone project, aimed at
understanding how our Milky Way galaxy was formed and how it has
evolved into what it is today. Gaia was selected by ESA in 2001 and
launched on 19 December 2013 from Kourou (French Guiana) with a
Soyuz-Fregat rocket. It was put on a Lissajous orbit around L2, 1.5
million km away from the Sun, for an expected five year lifetime of
operations, with a possible extension of one year.

Gaia has been designed to measure star positions with extreme accuracy
(see \cite{per01}). It will perform an all sky survey complete to
$G\sim 20.7$, corresponding to $V\sim 20-22$ mag for blue and red
objects respectively, obtaining astrometric data with micro arcsec
($\mu$as) accuracy for more than a billion stars, as well as optical
spectrophotometry for most of them, and medium resolution spectroscopy
at the Ca II triplet (847-870 nm) for several million objects brighter
than 16th mag. It is the successor to the Hipparcos mission, the first
precision-astrometry space mission, which produced a catalog of
positions, parallaxes and proper motions for about 100,000 stars. Gaia
will take this further by targeting $10^4$ more stars across the Milky
Way, and with an astrometric accuracy
$\sim100$ times better. It will provide in particular one-percent
accurate distance for 10 million stars (while it is the case for only
seven hundreds stars today).

\subsubsection{Payload and instruments}

\begin{figure}[htbp]
  \centering
  \includegraphics[width=0.8\textwidth]{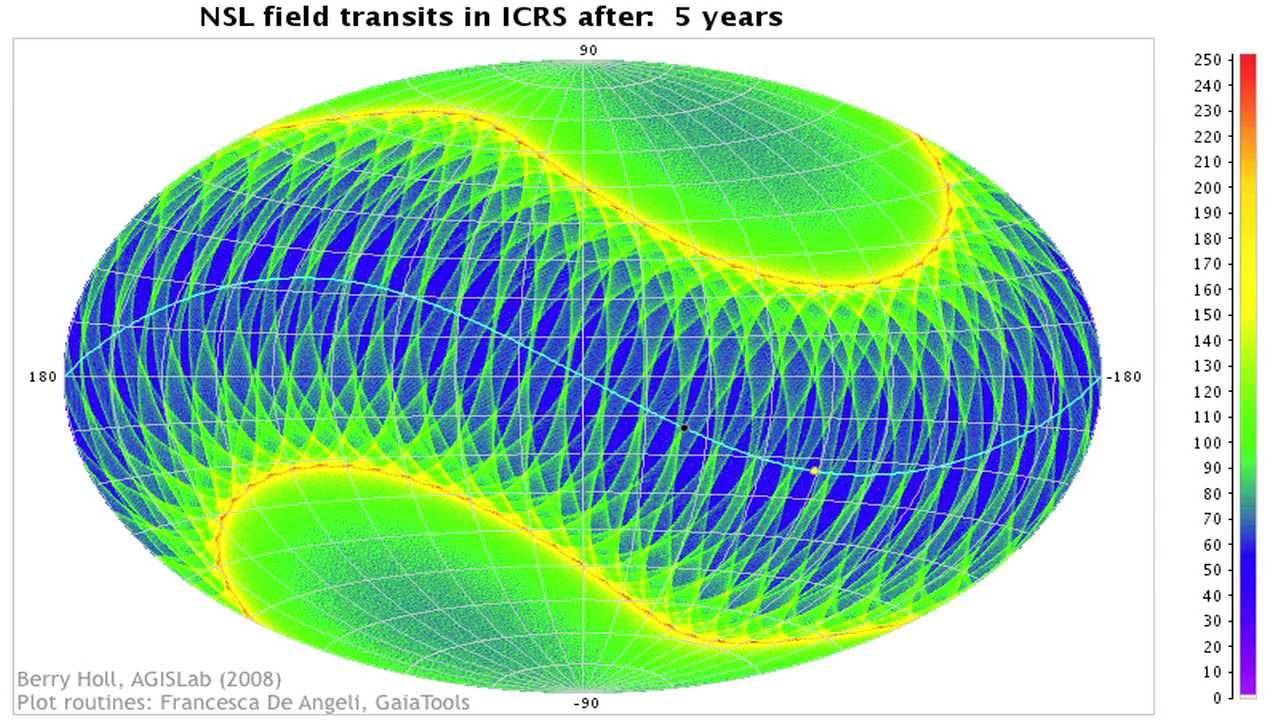}
  \caption{Gaia nominal skying law. The color scale indicates the
    number of Gaia visits on the sky in equatorial (ICRS) coordinates
    after 5 years. The light blue line corresponds to the ecliptic
    plane.}
  \label{gaia_nsl}
\end{figure}

Global astrometry requires repeated observations of the same area of
the sky from two lines of sight separated by a suitable and very
stable angle. For this purpose, Gaia's payload carries two identical
telescopes (each holding a 1.45m$\times$0.5m rectangular mirror),
pointing in two directions separated by a 106.5 degree basic angle and
merged into a common path in the focal plane. The satellite spins
around its axis, which is oriented 45 degrees away from the Sun, with
a period of 6 hours, and the spin axis has a precession motion around
the solar direction with a period of 63 days. The combination of these
two motions results in a quasi-regular time-sampling that allows to
scan the entire sky on average 70 times over the 5 year mission
lifetime, with more than 200 transits in a $\sim10$ degree strip at 45
degree latitude above or below the ecliptic plane, and as little as
$\sim$50 transits in other areas of the sky (see
Figure~\ref{gaia_nsl}).

The Gaia focal-plane assembly is the largest ever developed for a
space application, with 106 CCDs of 4500$\times$1966 pixels, yielding
a total of almost 1,000 million pixels of 10$\times$30 $\mu$m$^2$, and
a physical dimension of 1.0m $\times$ 0.4m. It serves five main
functions:
\begin{itemize}
\item the wave-front sensor (WFS) and basic-angle monitor (BAM);
\item the Sky Mapper (SM), autonomously detecting objects entering the
  fields of view and communicating details of the star transits to the
  subsequent CCDs;
\item the main Astrometric Field (AF), composed of 62 CCDs with a
  field of view of 40$\times$40 arcmin on the sky, devoted to
  astrometric measurements and $G$-band (330-1050 nm)
  photometry;
\item the Blue and Red Photometers (BP and RP), providing low
  resolution ($R<100$) slitless spectrophotometry for each object
  over the wavelength ranges 330-680 and 640-1000 nm,
  respectively, as well as integrated $G_{BP}$ and
  $G_{RP}$ magnitudes (and colour);
\item and the Radial-Velocity Spectrometer (RVS), registering slitless
  spectroscopy at the Ca II triplet (845-872 nm) with $R\sim 11,500$
  brighter than $G\sim16$ mag.
\end{itemize}

Measurements are made in time-delayed-integration (TDI) mode, reading the
CCDs at the same speed as the sources trail across the focal plane,
i.e., 1 arcmin/sec, corresponding to a crossing/reading time of 4.4 sec
per CCD. Since Gaia observes objects over a very wide range of apparent
magnitude, the CCDs must be capable of handling a wide dynamic signal
range. A number of features have been incorporated in the CCD design
in order to cope with bright stars. These include a large full-well
capacity, an anti-blooming drain, and TDI gates which effectively
reduce the integration time for bright objects.

The Gaia CCD detectors feature a pixel size of 10 $\mu$m (59
milli-arcsecond) in the scanning direction and the astrometric
instrument has been designed to cope with object densities up to
750,000 stars per square degree. In denser areas, only the brightest
stars are observed and the completeness limit will be brighter than
20th magnitude. 

\subsubsection{Science performance}

The commissioning phase was completed on 18 July 2014.  The overall
performance was assessed to be very good, and the detection efficiency
was improved so as to extend the bright end to $G\sim0$ mag through
detection algorithm improvements and employment of a special observing
mode. The faint limit was also extended down to $G = 20.7$ mag.
However, a few problems were identified: stray light (mostly due to
scattering from filaments on sunshield edges), contamination (thin ice
layer on optics), and some larger than expected variations of the
basic angle. These effects can be mitigated (e.g. by periodic
decontamination campaigns) or modelled, and actions have been taken to
this purpose. Updated performances can be found on the ESA website in
the Gaia performance
pages~\footnote{http://www.cosmos.esa.int/web/gaia/science-performance}.
They are summarized below.

\paragraph{Astrometry}

In order to get the five standard astrometric parameters, position
($\alpha$, $\delta$), parallax ($\pi$), and proper motion
($\mu_{\alpha}$, $\mu_{\delta}$) for all observed objects brighter
than $G = 20.7$ mag with micro-arcsec accuracy, all the available
$\sim 70$ observations per object gathered during Gaia's 5 year
lifetime will have to be combined in a single, global, and
self-consistent manner. While a daily data treatment is performed to
estimate source positions, satellite attitude and calibration
parameters at the level of sub-milli-arcsec accuracy, the Astrometric
Global Iterative Solution (AGIS) system is run about every 6 months on
an ever increasing data volume. AGIS treats the wanted source
parameters, the satellite’s attitude and calibration parameters as
unknowns and tries to find iteratively the best global match in a
least-square sense between all measurement data and an observational
model that is formulated in terms of these unknowns. This is the
reason why Gaia is sometimes referred to as a self-calibrating
mission.  Only single, non-variable stars which fit the standard
5-parameter astrometric model will take part in such a “primary” AGIS
cycle. For the remaining objects (binary, multiple systems, etc.) only
provisional values will be computed by AGIS in a subsequent
“secondary” cycle which optimizes source parameters using the attitude
and calibration solutions from the preceding primary cycle.

\begin{figure}[htbp]
  \centering
  \includegraphics[width=0.9\textwidth]{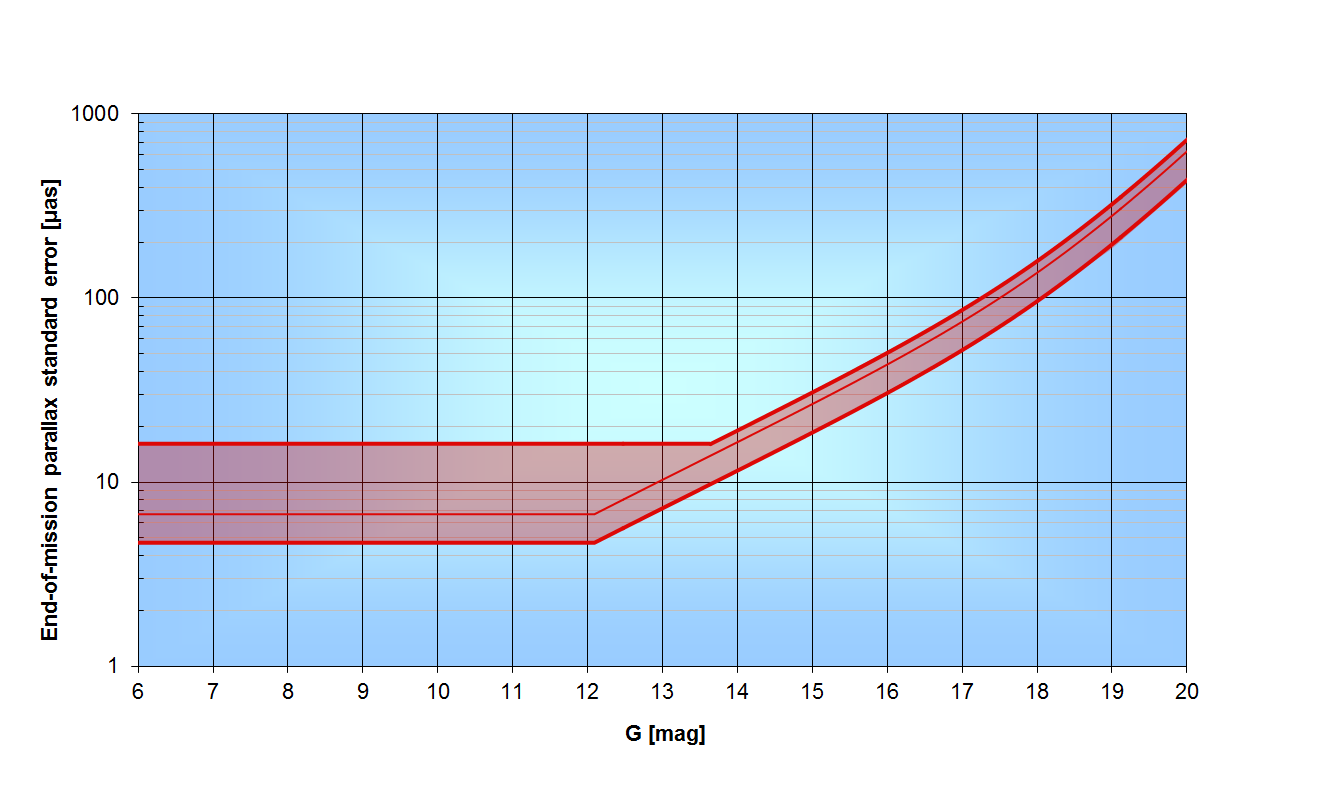}
  \caption{Sky-average end-of-mission parallax standard error, in
    $\mu$as, as function of Gaia $G$ magnitude for an unreddened G2V
    star ($V-I = 0.75$ mag, $V-G = 0.16$ mag). The upper and lower
    curves reflect the range resulting from varying the sky position,
    the $V-I$ colour index, and the bright-star (TDI-gate) observing
    strategy. Figure taken from the Gaia performance pages.}
  \label{sigma_pi}
\end{figure}
 
Taking into account all the known instrumental effects (except the
slow optics contamination by thin ice which is expected to remain small) under the
appropriate in-flight operating conditions plus a 20\% science
margin~\footnote{see
  http://www.cosmos.esa.int/web/gaia/science-performance for more
  details about this science margin}, the predicted end-of-mission
parallax standard error $\sigma_\pi$ is shown on
Figure~\ref{sigma_pi} for a single unreddened G2V star. All error
sources have been included as random variables with typical deviations
(as opposed to best-case or worst-case deviations). The behaviour of
the curves clearly shows two regimes: the bright star regime ($G<12$)
where performance is set by bright-star centroiding and calibration
errors, and the faint star regime ($G>12$) where photon-noise dominates.

Using scanning law simulations, the sky-averaged position and
proper-motion errors, $\sigma_0$ [$\mu$as] and $\sigma_\mu$
[$\mu$as/yr] can be deduced from $\sigma_\pi$  [$\mu$as] using the
following relations:
\begin{eqnarray*}
  \sigma_0 = 0.743 \,\,\sigma_\pi \\
  \sigma_\mu = 0.526 \,\,\sigma_\pi
\end{eqnarray*}

\begin{figure}[htbp]
  \centering
  \includegraphics[width=0.9\textwidth]{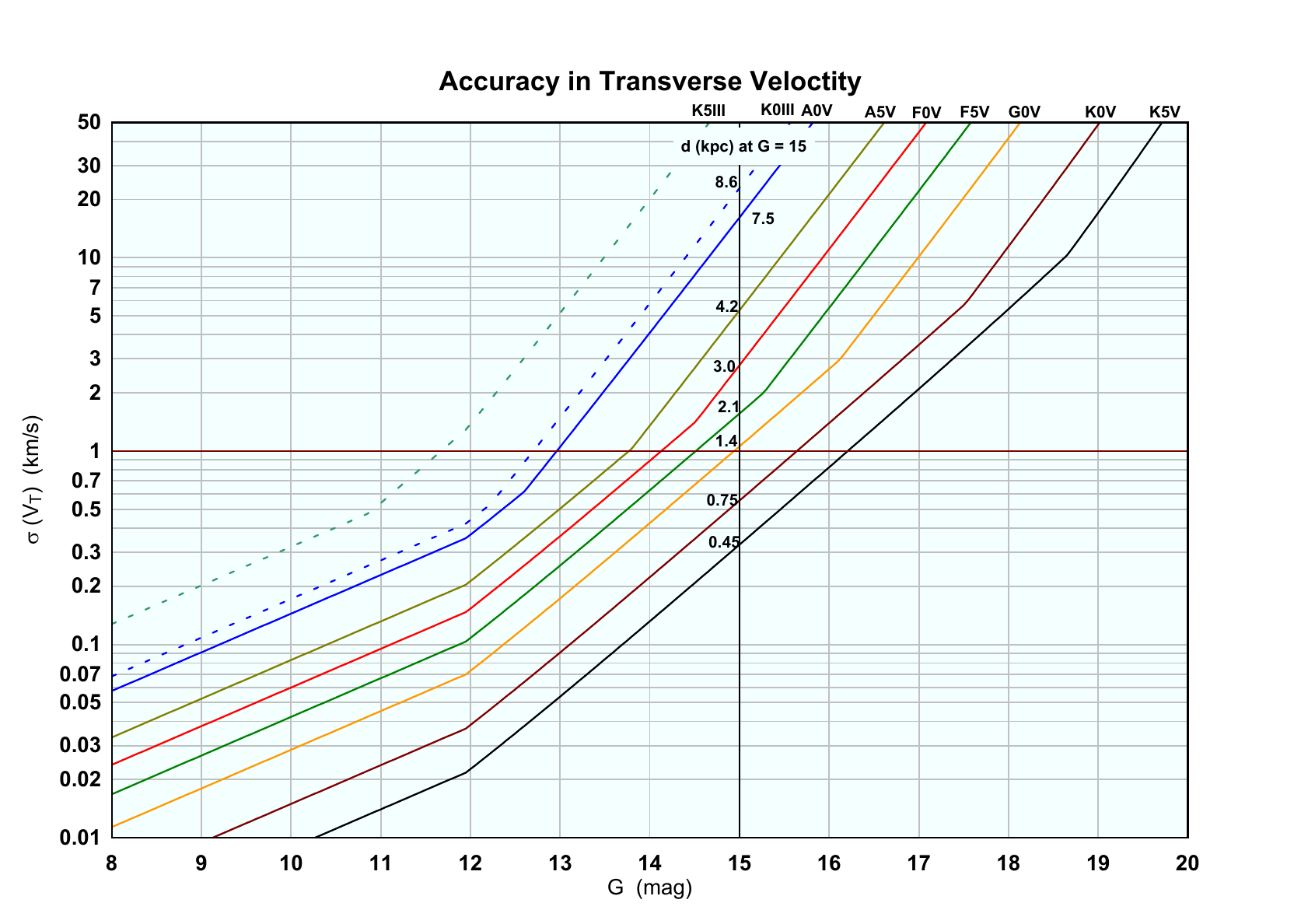}
  \caption{Expected performance of Gaia in the determination of the
    transverse velocity for a sample of representative stars, as a
    function of the magnitude in the Gaia broad band (the $G$
    magnitude). The uncertainties are determined from a combination of
    the uncertainties in the annual proper motion and parallax, as a
    function of the $G$ magnitude (from F. Mignard,
    GAIA-CA-TN-OCA-FM-048-1 technical note).}
  \label{vtan}
\end{figure}

Knowing that a star transverse velocity in km/s is given by $V_T=4.74
\mu/\pi$, with $\mu$ the star annual proper motion and $\pi$ its
parallax, the uncertainty on the transverse velocity $\sigma_{V_T}$
can easily be derived from $\sigma_\pi$ and $\sigma_\mu$. It is shown
in Figure~\ref{vtan} as a function of the $G$-band magnitude. In this
diagram, slope discontinuities appear first at $G = 12$, resulting
from the constant bright-star parallax accuracy adopted for $G < 12$,
and then at different magnitudes when the distance is equal to 2.4
kpc, i.e., when the Galactic shearing takes over on the random velocity
dispersion.

\paragraph{Photometry}

Gaia's photometry comprises: broad-band $G$-band fluxes obtained from
the astrometric instrument, and low-resolution spectro-photometry
obtained from the Blue (330-680 nm) and Red (640-1050) Photometers (BP
and RP). The wavelength coverage of the astrometric instrument
defining the $G$-band is $\sim$330-1050 nm. The Sloan/Cousins/Johnson
magnitude/colour transformations presented in \cite{jor10} allow
estimation of the broad-band Gaia $G$-magnitude as a function of $V$
and $V-I_C$. The spectral dispersion of the photometric instrument is
a function of wavelength and varies from 3 to 27 nm/pixel in BP and
from 7 to 15 nm/pixel in RP. Whereas the $G$-band data are
particularly useful for stellar variability studies, the BP/RP spectra
allow to classify sources and to determine their astrophysical
parameters, such as interstellar extinction, surface gravity,
effective temperature, etc. \cite{bj13} have shown that for FGKM stars
with magnitude $G = 15$ (i.e., $V\sim15-17$ depending on the source
colour), the effective temperature can be estimated to within
$\sim100$ K, the extinction $A_V$ to $\sim0.1$ mag, the surface
gravity to 0.25 dex, and the metallicity [Fe/H] to 0.2 dex, using just
the BP/RP spectro-photometry. They found that the accuracy of these
estimates depends on the $G$ magnitude and on the value of the stellar
parameters themselves, and may vary by a factor of more than two up or
down over this parameter range. These performances are based on
pre-launch simulations and would need to be confirmed on real data.

\paragraph{Spectroscopy}

Among the myriad of data collected by the ESA Gaia satellite, about
150 million high-resolution spectra will be delivered by the Radial Velocity
Spectrometer (RVS) for stars as faint as $G_{RVS}\sim16$ mag. 

\begin{figure}[htbp]
  \centering
  \includegraphics[width=0.9\textwidth]{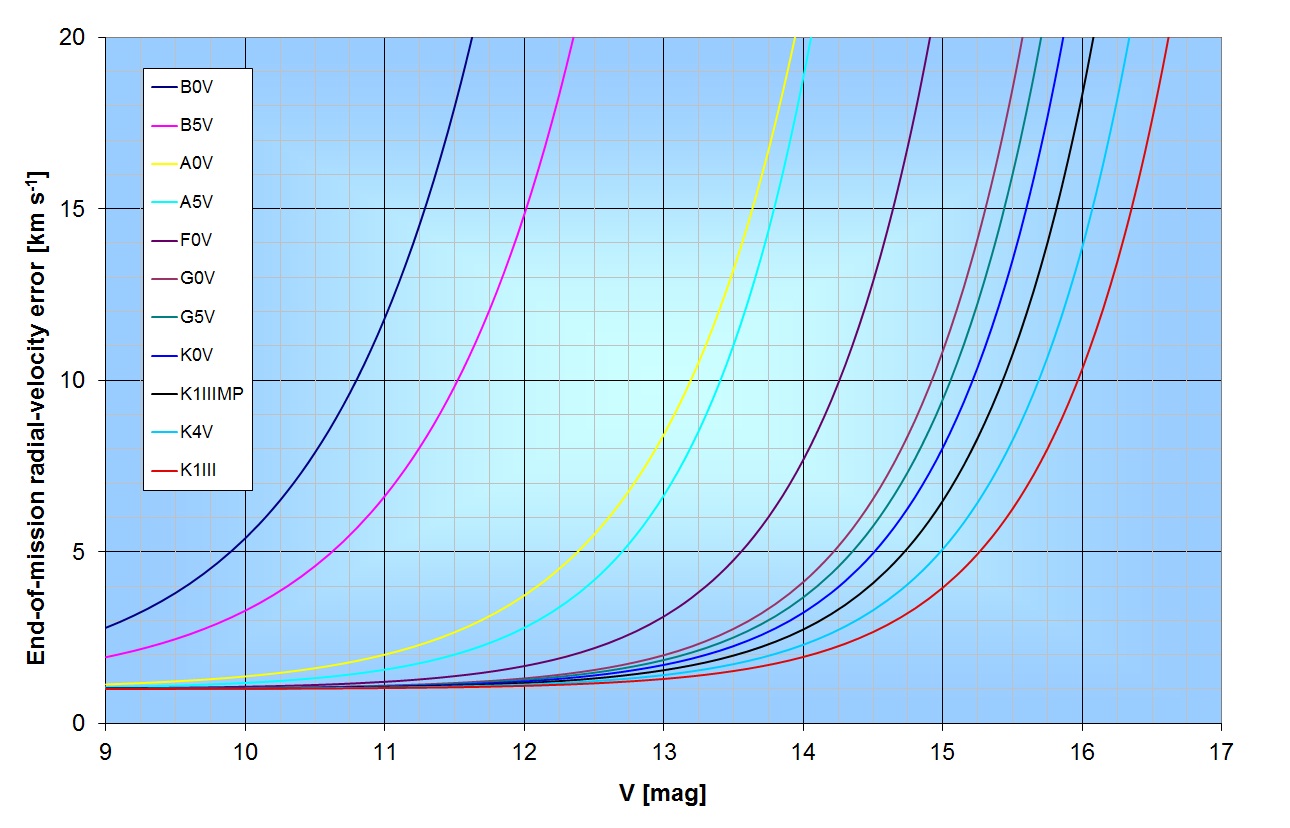}
  \caption{Expected RVS performance for a sample of representative
    stars, as a function of their $V$ magnitude. Plot available in the
    Gaia Performance Pages.}
  \label{RVSperf}
\end{figure}

Radial velocities are the main product of the RVS, allowing to
complete the five standard astrometric parameters obtained by the
Astrometric Field (AF) and obtain a 6-D view (in the position and
velocity phase-space) of the galactic stellar population.  The
calculation methodology prescribes that a single end-of-mission
composite spectrum is reconstructed by co-addition of all spectra
collected during the five-year mission, and an average radial velocity
is then extracted by cross correlation with a template
spectrum. Typical end-of-mission errors are 1 km/s at $V\sim7.5$,
12.3, and 15km/s at $V\sim11.3$, 15.2 for a B1V and G2V star
respectively (see Figure~\ref{RVSperf}). Whenever possible, an {\it a
  posteriori} on-ground data analysis will be done to derive also
single epoch radial velocities.

By comparison to the expected performance in transverse velocity
plotted in Figure~\ref{vtan}, one can see that globally the radial
velocity accuracy for G to K stars around $V = 15 - 16$ (corresponding
to the core of the Gaia mision), is smaller but only by a factor of a
few. For fainter stars, the RVS performance degrades much faster than
the astrometric determination of the transverse velocity and at the
faint end, there is no radial velocity measurement at all. For stars brighter
than $V\sim13$, the RVS performance remains rather constant, close to
1 km/s, while the expected transverse velocity accuracy is much
better, typically by one order of magnitude. Complementary ground
based observations are thus necessary to improve the radial velocity
accuracy at the level of the transverse velocity, a desirable feature
for kinematic studies.

In addition to radial velocity measurements, a specific stellar
characterization will be performed on single star RVS spectra with high
enough signal-to-noise ratio ($G_{RVS}<14.5$ mag), and some individual
chemical abundances will be estimated for the brightest targets
($G_{RVS}<11$ mag, $\sim2$ million stars). A recent study from
\cite{rb16} indicate that stars brighter than $G_{RVS}\sim12.5$ are
efficiently characterized, with reliable estimations of
[$\alpha$/Fe]. Typical internal errors for FGK stars are around 40 K
in T$_{eff}$, 0.10 dex in log(g), 0.04 dex in [M/H], and 0.03 dex in
[$\alpha$/Fe] at $G_{RVS} = 10.3$. They degrade to 155 K in T$_{eff}$,
0.15 dex in log(g), 0.10 dex in [M/H], and 0.1 dex in [$\alpha$/Fe] at
$G_{RVS} \sim 12$. Similar accuracies in T$_{eff}$ and [M/H] are found
for A-type stars, while the log(g) derivation is slightly more
accurate. For the faintest stars, with $G_{RVS}> 13-14$, a T$_{eff}$
estimate from the BP/RP spectrophotometry is necessary to allow a good
final RVS-based characterization.

\subsubsection{Gaia releases}

The Gaia mission is designed to operate for five years, with the possibility
of 12 to 18 months extension. The data have no proprietary rights and
will be released at planned intervals over the mission lifetime, after
a proper validation and with increasing details in the data
released. The release of the final catalog is planned for 3 years after the end of
the mission, i.e., around 2022 or 2023 (see the Gaia data release
page for updates).

The first data release is scheduled for the end of summer 2016, and
will contain positions and $G$-band photometry for single stars from
$\sim$90\% of the sky, and data from the ecliptic pole regions that
were scanned more often during commissioning. It will also deliver
positions, parallaxes and proper motions for a fraction of the 2.5
million Tycho-2 stars (\cite{hog00}). These astrometric data will be
obtained by combining the positions from the Tycho-2 Catalogue with
the first year of Gaia data for a joint Tycho-Gaia astrometric
solution (TGAS, \cite{mic15}; see also the Gaia webpage).  Thanks to
the $\sim$24 yr time baseline, these data will have sub-mas
accuracy. The second data release is currently planned for summer
2017, to deliver radial velocities for non-variable bright stars,
two-band (BP/RP) photometry, and full astrometric data where available
(90\% of the sky, only single stars). Two more data releases are
planned around 2018-19.  Each release updates the previous ones and
should contain significant new additions, including astrophysical
parameters and classification, mean radial velocities, orbital
solutions for binaries, variable star classification, and solar system
results. The final catalog should be consisting of:
\begin{itemize}
\item Full astrometric, photometric, and radial-velocity catalogues;
\item All available variable-star and non-single-star solutions;
\item Source classifications and astrophysical parameters (derived
  from BP/RP, RVS, and astrometry) for stars, unresolved binaries,
  galaxies, and quasars;
\item An exo-planet list;
\item All epoch data for all sources;
\item All ground-based observations made for data-processing purposes.
\end{itemize}

\subsection{Clusters and Gaia}

Accurate parallaxes and proper motions provided by the Gaia mission
will bring us into a new era of open cluster and association research.
Gaia will measure parallaxes and thus distances of {\it individual}
stars with a precision better than 1\% if closer than $\sim1$ kpc and
better than 10\% otherwise. Unprecedented accuracies are expected for
proper motions, yielding a precision on {\it individual} tangential
velocities of the order of 10 m/s and below for very nearby clusters
($<100$ pc), and better than 1 km/s for FGK stars in clusters up to 1
kpc and up to larger distances for bright O/B stars.  Gaia will also
provide good photometric information for all objects observed
astrometrically, as well as radial velocities and spectroscopic
characterization for the 15\% brighter.

Gaia will thus deliver a dataset, for both known and possibly newly
discovered clusters and associations, that will have a huge impact on
a large variety of topics on young cluster research. A
non-comprehensive list is given below.

\paragraph{Dynamics of open clusters and associations: formation and
  evolution}

Cluster origin theories range from highly dynamic models
(\cite{bon03}) to quasi equilibrium and/or slow contraction scenarios
(\cite{tan06}). The two scenarios lead to different initial cluster
structure and kinematics. For example, shallow or absent velocity
gradients are expected if clusters form in a quasi-equilibrium state,
while strong gradients should be present if clusters form in
sub-virial conditions (e.g., \cite{pro09}). In particular, the
properties of ejected stars that will be kinematically recovered by
Gaia will be vital for constraining the dynamics of star forming
regions. For example their age distribution will provide constraints
on the timescale of star formation, distinguishing between slow
(\cite{kru05}) and rapid (\cite{bal99}) models of star
formation. Their spatial and kinematics properties at birth will be
obtained by tracing them back to their natal regions, which will
reveal the initial structure and density of star forming regions, as
well as any possible cluster rotation and kinematic substructures.

Gaia will also provide an opportunity to address questions such as
whether OB associations are expanding star clusters or whether they
form with low densities (\cite{wri14}), and whether massive stars can
form in isolated or low-density environments, or whether they have
been dynamically ejected from dense regions (\cite{bre12}). It should
in particular be possible to resolve the origins of OB associations by
tracing back the motions of their members to see if they represent an
expanded star cluster. For simple systems this may be possible with
early Gaia data, though more complex arrangements of multiple
expanding star clusters will require parallaxes and more precise
proper motions from later data releases.

Moreover, a common model of star formation suggests that most stars
are born in clusters, but that the majority of these clusters become
gravitationally unbound when feedback disperses the gas left over from
star formation and get disrupted. This theory of cluster disruption,
often referred to as residual gas expulsion, predicts that a large
fraction of young and non-embedded star clusters should be supervirial
and in the process of expanding.  Kinematic Gaia data will be
fundamental to address this issue as it will provide a measurement of
the dynamical state of a cluster or association, either using proper
motions and/or radial velocities (after correction for the broadening
of the velocity dispersion due to binaries). Such studies will
provide an indication of whether the regions are in virial equilibrium
or in the process of collapsing or expanding, and thus constrain any
possible effect of residual gas removal.

Also, depending on several factors such as initial conditions
themselves, SF efficiency, tidal interaction with the Galaxy, clusters
can undergo infant mortality on very short timescales ( $< 10 - 30$
Myr; e.g., \cite{fal06}) or experience longer term dissolution
(\cite{dgp07}).  In order to understand cluster evolution and to put
constraints on those parameters which essentially determine whether
the cluster will disrupt into the field or not, realistic simulations
are needed. While progress has been made during the past years in the
development of these (e.g., \cite{tas08}; \cite{cap10}), the vital
missing element is a large body of observations to be compared with
these models, which will be provided by Gaia. N-body numerical
simulations of cluster dynamical evolution make indeed many
predictions that will be tested using kinematics from Gaia, including
how the velocity dispersions of stars vary as a function of stellar
mass and environment, the evolution of mass segregation, the expansion
and evaporation of star clusters, and their distortion due to the
galactic tidal field that can be studied using proper motions.

\paragraph{Stellar structure and evolution}

Gaia will determine accurate distances of a few hundred star clusters
in our galaxy for which complete and accurate lists of members will be
available. This will finally make fully accessible the unique
potential of open clusters to empirically resolve the
Hertzsprung-Russell diagram (HRD) and to put tight constraints on the
so far still quite limited calibrations of stellar evolutionary models
for Pop. I stars, from the pre-main sequence (PMS) phase up to the
latest evolutionary stages. Indeed, input physics (opacities, EOS,
treatment of convection, mixing, rotation, magnetic fields, He
abundances) that significantly influences the stellar luminosity,
lifetime and burning-phase duration, still need to be calibrated
(e.g., \cite{nay09}; \cite{sod10}). Reliable stellar masses and ages
can be determined only once the models are appropriately adjusted to
represent the data for a wide range of open clusters of different ages
and compositions. This is vital for all areas of stellar and galactic
astrophysics.  Individual distances and photometry provided by Gaia
for nearby open clusters and for large samples of members, covering
a wide range of spectral types and masses, will yield exquisitely
precise colour magnitude diagrams (CMDs) which will be translated into
a set of observed isochrones. A comparison of these isochrones with
those obtained through modelling will enable us to adjust the models
and learn more about stellar evolution, stellar structure and stellar
atmospheres.

\paragraph{The population and evolution of the Galactic thin disk}

Key questions are how clusters populate the Milky Way thin disk and
what they tell us about its formation and evolution. To answer the
first question, we need the data for many OCs at a wide range of
Galactocentric radii ($R_{GC}$), and with a large range of ages. The
unprecedented accuracy of the astrometric Gaia data will allow us to
identify stars sharing the same kinematics, yielding the discovery of
new clusters and associations up to 5kpc away from the Sun. High
resolution spectroscopic data will then be crucial in order to check
that the new groups are not just streams but that their members share
the same origin, i.e., have the same radial velocity, the same age,
and the same abundance pattern.

The very young clusters trace the areas of star formation, while older
clusters trace the dispersion. The survival rate of OCs as a function
of age can be used to trace the history of field star populations
(e.g., \cite{deg10}).  OCs of different ages and $R_{GC}$ also reveal
information on the radial abundance gradient in the thin disk and on
its evolution (e.g., \cite{ran05}; \cite{bt06}). This is a key input
to models of formation and evolution of the disk as metallicity may
affect the SF efficiency and the shape of the initial mass function,
which will be much better constrained with Gaia. Thanks to the very
precise proper motion measurements, ejected cluster members will be
recovered by tracing them back to their natal regions, providing a
complete census down to the substellar regime for the closest and/or
less extincted ones. This will allow us to be much less sensitive to
uncompleteness and to the cluster dynamical evolution when inferring
its luminosity function and IMF.

The abundance gradient at different ages is also essential for
assessing the radial streaming of gas (e.g., \cite{sb09}), the
co-rotation resonance, the influence of the bulge, the nature of
infall from the halo, and the recent history of our Galaxy spiral
pattern as well as the role of possible mergers (e.g., \cite{mag09}).
Current estimates of the gradient suffer from low number statistics,
lack of homogeneity, of samples at key positions/ages, and of orbit
determinations (e.g., \cite{mag10}).  Key new information that will be
provided by Gaia are accurate distances from the Sun, allowing to
reduce the uncertainties in Galactocentric radii, and proper
motions. Additional high resolution spectroscopy will then be necessary
for accurate abundance determination of several elements in order to
put constraints on previous generations of polluters and to
reconstruct the chemical evolution of the disk.



\subsection{Complementary studies and future facilities}

Despite its extraordinary accuracy in parallaxes, positions and proper
motions, one of the main limitation of Gaia is its spectroscopic
capability. Not only is the limiting distance for radial velocity
measurements smaller than for photometry and astrometry (only stars
brighter than $G=16$ mag will have a spectroscopic observation), but
the expected accuracy (1 km/s at best) is much below that expected for
the transverse velocity, especially for the brightest objects
($G<12-13$). Moreover, abundance determinations will be possible only
for stars brighter than $G=12$.

The other strong limitation of Gaia is that it observes in the
optical. As such, it is blind to faint very red objects and sources in extincted
regions. Looking for brown dwarfs in star forming regions will thus
remain a challenge and will only be possible in close and
not too embedded regions.

Complementary data are thus required to obtain more accurate ($<1$
km/s) radial velocities and get proper motions at fainter magnitudes
and in extincted regions, as well as to measure abundance
patterns. Below is a non-exhaustive list of existing and future
surveys that will greatly complement Gaia data. P.I. observations may
however still be required to obtain high resolution spectroscopic data
and/or deeper and NIR proper motions for specific regions.

\subsubsection{On-ground spectroscopic studies}

\paragraph{RAVE}

RAVE (RAdial Velocity Experiment) is a multi-fiber spectroscopic
astronomical survey of stars in the Milky Way using the 1.2-m UK
Schmidt Telescope of the Australian Astronomical Observatory
(AAO).

From 2003 to 2013 (end of observation campaign), RAVE has covered
20,000 square degrees of the sky, obtaining spectra for about 500,000
stars in the magnitude range $8 < I < 12$ mag, and with a distance up
to $\sim3$ kpc from the Sun. The covered spectral region contains the
infrared Calcium triplet and is similar to the wavelength range chosen
for Gaia's Radial Velocity Spectrometer. The effective resolution of
$R \sim 7,000$ enables radial velocity measurements with a median
precision better than 1.5 km/s, as well as good precision atmospheric
parameters (effective temperature and surface gravity) and chemical
abundances for the surveyed stars.

The RAVE database is public and can be accessed from the RAVE website
(https://www.rave-survey.org/).

\paragraph{GALAH}

The GALactic Archeology with HERMES
project started in 2014 (see https://galah-survey.org/home).  Observations
are carried out with HERMES, a multi-fibre spectrograph operating on
the 3.9 m Anglo-Australian telescope, in four wavelength ranges
in the visible covering up to 30 different elements with a resolution
$R\sim28,000$.  The instrument uses the existing 2dF optical fibre
positioner to place the 400 fibres over a two-degree field of
view. The survey is limited to stars brighter than $V =14$.  Even so,
it is providing a very important ``all-sky'' (below $+10$ deg
declination) complement to Gaia, delivering high quality elemental
abundances for almost a million stars.  GALAH uses open and
globular clusters for calibrating purposes.

\paragraph{Gaia-ESO Survey (GES)}

The public spectroscopic Gaia-ESO survey, is targeting more than 100,000
stars using FLAMES at the VLT, and covers the bulge, thick and thin
disks and halo components, as well as a significant sample of 100 open
clusters of all ages and masses.

The survey uses the GIRAFFE and UVES spectrographs to measure detailed
abundances for at least 12 elements (Na, Mg, Si, Ca, Ti, V, Cr, Mn,
Fe, Co, Sr, Zr, Ba) in up to 10,000 field stars with $V < 15$ mag and
for several additional elements (including Li) for more metal-rich
cluster stars. The radial velocity precision for this sample is
0.1 to 5 km/s, depending on target, and about 0.2 km/s for open
cluster stars (\cite{jac15}).

Public Gaia-ESO Survey Science Archive can be found at
http://ges.roe.ac.uk/.

\paragraph{APOGEE}

The Apache Point Observatory Galactic Evolution Experiment (APOGEE)
was one of the programs of SDSS-III. The survey consists of
high-resolution ($R\sim22,500$), high signal-to-noise (S/N$>100$)
near-infrared spectra (wavelength range from 1.51 to 1.70 $\mu$m) of
mainly 100,000 red giant stars across the full range of the Galactic bulge,
bar, disk, and halo using the 2.5m Sloan telescope. APOGEE North and
South (also known as APOGEE-2) is now extending this survey in the
southern hemisphere using also the 2.5m du Pont Telescope at Las
Campanas Observatory in Chile, and targeting $\sim300,000$ stars down
to $H\sim13$. The radial velocity precision is around 0.1 km/s and the
abundance precision is 0.1 dex for more than 15 elements.

Open clusters have also been targeted with APOGEE and APOGEE-2
yielding the Open Cluster Chemical Abundances and Mapping (OCCAM)
survey (\cite{fri13}), and the INfrared Survey of Young Nebulous
Clusters (IN-SYNC, e.g. \cite{cot14}).

\subsubsection{On-ground proper motion surveys}

\paragraph{UKIDSS/GCS}

The Galactic Cluster Survey (GCS) was one of the five UKIDSS surveys
dedicated to open clusters. The GCS imaged 10 open star clusters and
star formation associations, covering a total area of 1400 sq. degs,
in $JHK$ to a depth $K=18.7$. The GCS has been completed in two
stages. The first stage was a single pass in $JHK$ over 1400 sq. deg.
while the second stage was an additional pass in $K$ over the same
1400 sq. deg., for proper motion measurements, with a time baseline
of 2 to 7 years and a typical accuracy of 2-6 mas/yr (e.g. \cite{lod12}).

The UKIDSS data can be accessed from the WFCAM Science Archive (WSA,
http://wsa.roe.ac.uk/).

\paragraph{Pann-STARRS}

Pan-STARRS (Panoramic Survey Telescope \& Rapid Response
System, http://pan-starrs.ifa.hawaii.edu/public/) is a
wide-field imaging facility developed at the University of Hawaii's
Institute for Astronomy and located at Haleakala. The first telescope
(PS1) is a 1.8 meter telescope with a 3.2 degree diameter field of
view and is equipped with a 1.4 Gigapixel Camera. The second telescope
PS2 should become operational soon.

The PS1 3$\pi$ survey produced a catalog of more than three
billion objects north of $-30$ declination with astrometric precision
down to 10 milliarcseconds per observation. This excellent calibration
and the multiple observations of each point of the sky over a five
year internal baseline (60 epochs for each stars, twelve each in 5
filters) allow proper motions as small as 1-2 mas/yr to be
measured. The observational schedule has been designed to optimize
parallax measurements for red objects. The final parallax and proper
motion catalog will enable searches for and classification of stars in
the local solar neighborhood, with proper motion searches extending to
200 parsecs.

All the data of PS1 surveys will become public at the end of the PS1
science mission.

\paragraph{DANCe}

\begin{figure}[htbp]
  \centering
  \includegraphics[width=0.9\textwidth]{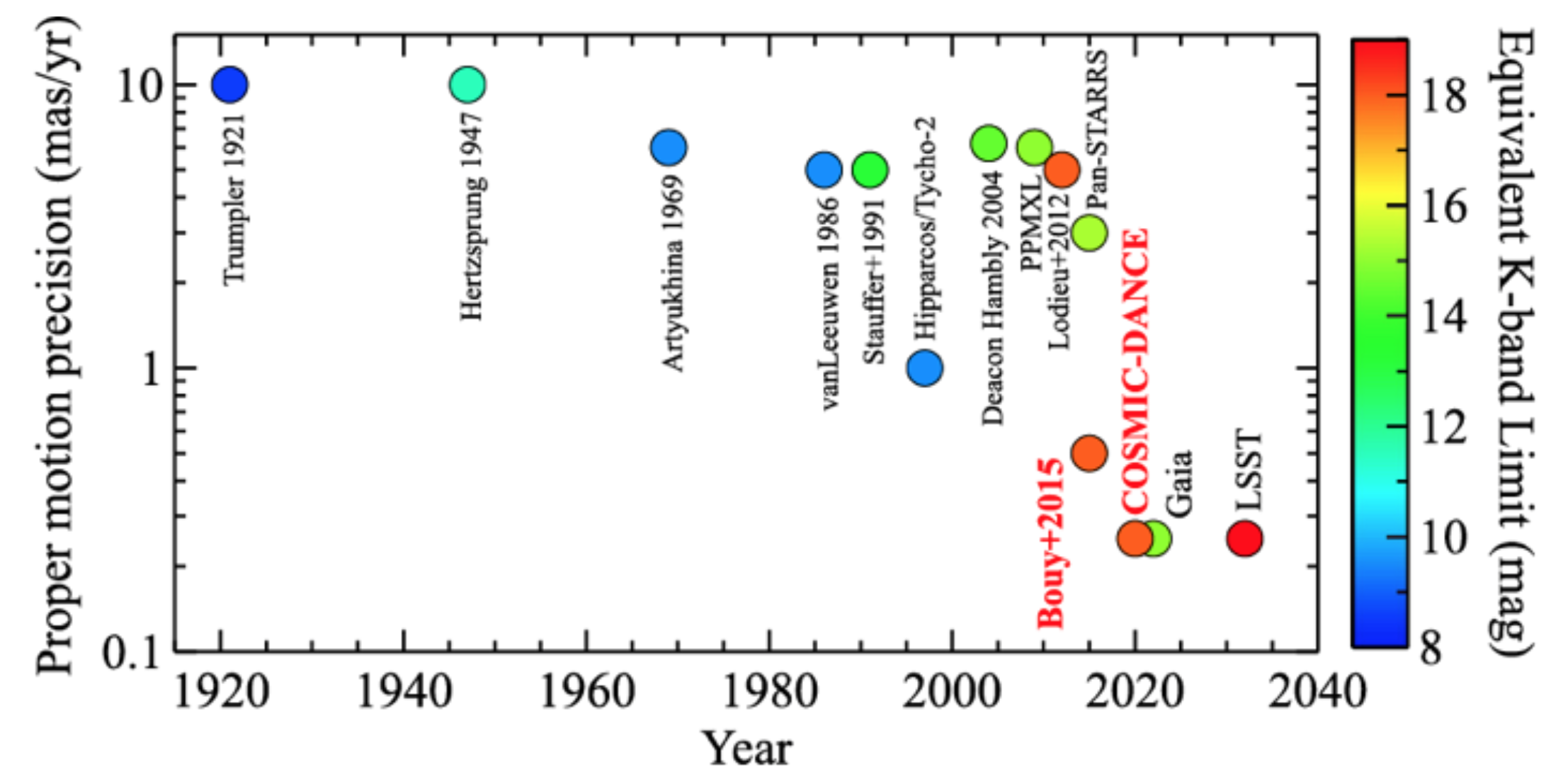}
  \caption{100 years of proper motion studies in the
    Pleiades. COSMIC-DANCE can improve the precision and depth by an
    order of magnitude, and remain unsurpassed until LSST ends in
    2032. For optical surveys, the equivalent K-band limit was
    estimated using typical colours of Pleiades members. Courtesy from
    H. Bouy.}
  \label{dance}
\end{figure}

The Dynamical Analysis of Nearby Clusters (DANCe) project
(http://www.project-dance.com) is an ambitious ground based survey
aiming at measuring precise proper motions for millions of faint stars
in young nearby clusters, taking advantage of archival wide-field
surveys. Using the Pleiades as a test bench, \cite{bou13} have
developed a powerful collection of softwares for the processing of
large amounts of wide-field images from various instruments to provide
astrometric (proper motions) and photometric catalogs. 17,000 archival
and new images obtained over 15 years and covering 80 deg$^2$ in the
Pleiades have been combined to reach a final accuracy between 0.3 and
1 mas/yr for all the stars in the magnitude range $i=12-23$ (see
Figure~\ref{dance}). DANCe is four magnitudes deeper than Gaia,
reaching the planetary mass domain and piercing through the dust
obscured young clusters inaccessible to Gaia's optical sensors. The
extension of this pilot study is named COSMIC-DANCE and will cover 20
young ($<120$ Myr) and nearby ($<410$ pc) clusters and associations,
including Gaia-ESO targets when possible.

\subsubsection{Future facilities}

Several optical/near-infrared high-resolution multi-object
spectrographs are currently under development in order to be able to
measure precise radial velocities and abundance patterns to follow-up
Gaia.

WEAVE (http://www.ing.iac.es/weave/) is a new multi-object survey
spectrograph for the 4.2m William Herschel Telescope at the
Observatorio del Roque de los Muchachos, on La Palma in the Canary
Islands. It will allow to take optical spectra of up to ~1000 targets
over a two-degree field of view in a single exposure. The fibre-fed
spectrograph comprises two arms, one optimised for the blue and one
for the red, and offers observations at $R\sim5000$ over the full
370-1000nm wavelength range in a single exposure, or a high resolution
mode with limited coverage in each arm at $R\sim20,000$. The system
integration has now started and the first light at the telescope is
expected early 2018.

MOONS at the VLT (http://www.roe.ac.uk/~ciras/MOONS/VLT-MOONS.html)
should be available one year after (in 2019). The baseline design
consists of $\sim$1000 fibers deployable over a field of view of
$\sim$500 square arcmin. The total wavelength coverage is
0.6$\mu$m-1.8$\mu$m with two resolution modes. In the medium
resolution mode ($R\sim4000-6000$) the entire wavelength range is
observed simultaneously, while the high resolution mode covers
simultaneously three selected spectral regions: one around the CaII
triplet (at $R\sim9,000$) to measure radial velocities, and two
regions at $R\sim20,000$ (one in the $J$-band and one in the $H$-band)
for detailed measurements of chemical abundances.

4MOST (https://www.4most.eu/) is a fibre-fed spectroscopic survey
facility on the 4m ESO/VISTA telescope that will be able to
simultaneously obtain spectra of $\sim$2400 objects distributed over a
4 square degrees field of view.  To reach maximum impact, it will
operate continuously for an initial five-year Public Survey delivering
spectra for $>25$ million objects over $>15,000$ square degrees at
$R\sim5000$ and $\sim2$ million targets at $R\sim20,000$.  4MOST is
currently in its Preliminary Design Phase with an expected start of
science operations in 2021.

Finally, the Maunakea Spectroscopic Explorer (MSE,
http://mse.cfht.hawaii.edu/) is a project to build a 10-meter class
telescope replacing the CFHT, and fully devoted to spectroscopy. The
single dedicated MSE instrument is a fibre-fed multi-object
spectrograph, capable of gathering light from up to 3200 simultaneous
objects, each being measured with spectral resolution from $R\sim2000$
over the entire 370-1300nm wavelength range to about 20,000 over a
restricted range. MSE would be a survey instrument.

Concerning future photometric facilities for proper motions
measurements, the Large Synoptic Survey Telescope (LSST, see
http://www.lsst.org/) is probably the most awaited one. It will
produce a 6-band (corresponding to the $ugrizy$ filters from 0.3-1.1
micron) wide-field deep astronomical survey of over 20,000 square
degrees of the southern sky (south of $+10$ deg declination) down to
$r\sim28$ AB mag using an 8.4-meter ground-based telescope. Each patch
of sky will be visited about 1000 times in ten years thanks to the
very large field of view of the telescope (about 10 sq.deg.).  The
expected parallax and proper motion accuracy for a ten-year long
baseline is 0.6 mas and 0.2 mas/yr at $r\sim21$, and 2.9 mas and 1
mas/yr at $r\sim24$. This is fairly similar to Gaia performance for
faint objects, but the real advantage of the LSST is that it will be
much deeper (by almost 6 magnitudes) and is more sensitive to distant
and red or extincted objects. For example, proper motion of about 200
million F/G main-sequence stars will be detected out to a distance of
100 kpc; complete census of faint populations in nearby star forming
regions using color and variability selection will be obtained; deep
and highly accurate color-magnitude diagrams for over half of the
known globular clusters, including tangential velocities from proper
motion measurements, will be created; the Milky Way's halo and the
Local Group out to 400 kpc will be mapped. The LSST is currently being
built in Chile and full science operations for the ten-year survey
should begin in 2023.

Although not wide field, the JWST (with a launch planned for 2018) and
the E-ELT (expected for 2024) are also very much awaited. They will
particularly be vital to probe the core of extincted star forming
regions thanks to their unprecedented sensitivity in the
infrared. Moreover, the E-ELT will benefit from an exquisite spatial
resolution that will provide access to crowded/distant regions for the
first time and allow the characterization of very faint/multiple
objects. This will allow for example i) to obtain the subsolar census
of young massive clusters and more distant environments such as the
LMC/SMC in order to test for the IMF universality at high density
and/or low metallicity; ii) to measure their internal dynamics and
parallaxes from astrometry (precision $\sim50\mu$as) , and iii) to
probe short separation and very low mass ratio binaries in nearby
clusters.

Thanks to the above facility developments and the existing surveys, we
are now in a really good position to take full advantage of the wealth
of Gaia data. This will truely revolutionize our view of the formation
and evolution of clusters and associations, of the stars themselves
and of the Galactic thin disk as a whole as we will have access for
the first time to the 6 phase-space dimensions of the solar
neighborhood up to 1kpc (and then 100kpc with LSST), and to accurate
photometry and abundances.


\section{Conclusion}

It is clear that the Gaia mission is going to revolutionize our
understanding of the stellar cluster formation and evolution as well
as our view of the Milky Way. To fully exploit the immense amount of
data coming from both Gaia and complementary programmes one needs to be
prepared however. The data analysis will require new statistical tools
to efficiently study young stars and star clusters in 6-dimensional
phase space (three position and three velocity dimensions), as well as
folding in a wealth of multi-dimensional data.

Moreover, hybrid hydrodynamic and N-body simulations of star forming
regions are needed to link the collapse of molecular clouds with the
dynamical evolution of young embedded stellar clusters.  We will need
many simulations that cover a wide variety of initial and final
conditions as well as a number of new statistical diagnostics that
will allow us to quantitatively compare observations and simulations.
These diagnostics will have to be extensively tested against
simulations to make sure that they reliably trace the cluster
evolution and that they can be used to resolve the initial conditions.


\end{document}